\documentclass[aps,showpacs,preprintnumbers,nofootinbib,tighten]{revtex4}
\usepackage{graphicx}% Include figure files
\usepackage{dcolumn}% Align table columns on decimal point
\usepackage{bm}% bold math
\usepackage{amssymb,amsmath}%symbol 
\usepackage{epsfig}

\begin{document}

\title{\Large{Wakes in the quark-gluon plasma }} 

\author{Purnendu Chakraborty,}\email{purnendu.chakraborty@saha.ac.in}
\author{Munshi G. Mustafa}\email{munshigolam.mustafa@saha.ac.in}

\affiliation{Theory Group, Saha Institute of Nuclear Physics, 1/AF Bidhan Nagar,
Kolkata 700 064, India}

\author{Markus H. Thoma}\email{mthoma@ipp.mpg.de}

\affiliation{Max-Planck-Institut
f\"ur extraterrestrische Physik, P.O. Box 1312, 85741 Garching, Germany}

\vspace{0.2in}

\begin{abstract}
Using the high temperature approximation we study, within the linear 
response theory, the wake in the quark-gluon plasma by a fast parton owing to
dynamical screening in the space like region.  
When the parton moves with a speed less than the average speed of the 
plasmon, we find that the wake structure corresponds to a screening charge 
cloud traveling with the parton with one sign flip in the induced charge 
density resulting in a Lennard-Jones type potential in the outward flow with 
a short range repulsive and a long range attractive part. 
On the other hand if the parton moves with a speed higher than that of 
plasmon, the wake structure in the induced charge density is found to have
alternate sign flips and the wake potential in the outward flow oscillates 
analogous to Cerenkov like wave generation with a Mach cone structure trailing
the moving parton. The potential normal to the motion of the parton indicates 
a transverse flow in the system.  We also calculate the potential due to a 
color dipole and discuss consequences of possible new bound states and 
$J/\psi$ suppression in the quark-gluon plasma. 

\end{abstract}

\pacs{12.38.Mh,24.85.+p}

\maketitle

\vspace{0.2in}

\section{Introduction}
\label{sec_intro}
A plasma is a statistical system of charged particles which move randomly, 
interact with themselves and respond to external disturbances.
Therefore, it is capable of sustaining rich classes of physical phenomena.
Screening of charges, damping of plasma modes, and plasma oscillations are
important collective phenomena in plasma physics~\cite{Ichimaru73,Landau81}. 
A proper description of such phenomena may be obtained if we know how a 
plasma will respond macroscopically to a given external disturbance. 
The microscopic features of the particle interactions in the plasma are not 
completely lost in such a macroscopic description. But they can be
implemented in a response function through the ways in which the mutually
interacting particles adjust themselves to the external disturbance and 
the response function plays a crucial role in determining the properties 
of the plasma~\cite{Ichimaru73}. 
 
Soon after the discovery of Quantum Chromodynamics (QCD), it has been found 
that at high temperature $T$ the color charge is screened~\cite{Shuryak78}
and the corresponding phase of matter was named Quark-Gluon Plasma (QGP).
It is a special kind of plasma in which the electric charges are replaced by
the color charges of quarks and gluons, mediating the strong interaction
among them. Such a state of matter is expected to exist at extreme temperatures,
above $150$ MeV, or densities, above about $10$ times normal nuclear 
matter density. These conditions could be achieved in the early universe
for the first few microseconds after the Big Bang or in the interior 
of neutron stars. In accelerator experiments high-energy nucleus-nucleus
collisions are used to search for the QGP. These collisions create a hot and dense 
fireball, which might consist of a QGP in an early stage (less than about 
$10$fm/$c$)~\cite{Muller85}.  Since the masses
of the lightest quarks and of the actually massless gluons are much less
than the temperature
of the system, the QGP is an ultrarelativistic plasma. To achieve a
theoretical understanding of the QGP, methods from quantum field theory (QCD)
at finite temperature are adopted \cite{Kapusta,Lebellac}.
Perturbative QCD should work at
high temperatures far above the phase transition where the interaction between
the quarks and gluons becomes weak due to 
asymptotic freedom. An important quantity which can be derived in this way
is the polarization tensor describing the behavior of interacting
gluons in the QGP. The dielectric function is related to the polarization 
tensor and important properties of the QGP, such as the dispersion 
relation of collective plasma modes and their damping or the Debye screening 
of color charges in the QGP~\cite{Kajantie}.

Recent numerical lattice calculations have found that charmonium states 
remain bound at 
least up to $T\sim 2T_C$~\cite{Datta} and the behavior of temporal correlators 
in pseudoscalar and vector channels deviates significantly from the free behavior 
at $T\sim 3T_C$~\cite{Karsch}. These analysis suggested {\it that pseudoscalars
and vectors resonances may exist above $T_C$}. Also  robust results from Au+Au 
at the BNL Relativistic Heavy Ion Collider (RHIC) have shown collective 
effects known as radial~\cite{radial} and elliptic~\cite{elliptic,elliptic1} 
flows, and a suppression of high-$p_\bot$ hadron spectra~\cite{elliptic1,RHIC},
which could possibly indicate the quenching of light quark and gluon 
jets~\cite{Pluemer}. The hydrodynamical description~\cite{fluid} of the 
observed collective flow indicates that the matter produced at RHIC {\it 
behaves like a near-perfect fluid}. 
On the other hand, the amount of jet quenching might depend on the state of 
matter of the fireball, i.e., QGP or a hot hadron gas. There are extensive 
theoretical efforts in understanding the effect of the medium on jet 
quenching~\cite{Gyulassy,Salgado,Wang,Baier,Mueller,Munshi,Jane} using 
collisional as well as radiative energy loss since the high energy partons
traveling through a medium will lose energy owing to the interactions 
in the medium. 
%Hence {\it jet will be quenched}. 

Recently Ruppert and M\"uller~\cite{Ruppert} have studied the wake 
of the in-medium jet physics assuming a fast moving color charge particle with
high momentum in a QGP within the framework of the linear response theory, 
considering two different scenarios: 
{\sf (i) a weakly coupled QGP as described by Hard Thermal
Loop (HTL) perturbation theory and (ii) a QGP with the properties of a
quantum liquid}. In both cases the wake of the medium, i.e., QGP, has been 
observed through the induced charge and current densities due to dynamical 
screening but only in the latter case (ii) the wake exhibits 
a oscillation analogous to Cerenkov like radiation with a Mach cone 
structure. We point out here that their analysis within the HTL perturbation 
theory is not complete. 
%as they {\sf overlooked some important features of the 
%medium's response to the fast moving jet}. 
In this paper we revisit their analysis within the HTL perturbation 
theory extending our investigations of dynamical screening~\cite{Mustafa05}. 

In Ref.\cite{Mustafa05} we restricted ourselves to the outward flow of the
wake potential and to small particle velocities ($v \leq 0.8$). Also in the
two-body potential for describing parton-parton interaction and bound states
we did not include the magnetic interaction. Here we will show that both 
effects should not be neglected in the case of high-energy partons. 
Furthermore, in our earlier study we concentrated only on the screening 
aspect of the wake potential, whereas here we will discuss also the 
relationship to the energy loss and dynamic aspects (oscillations, flow).

In particular, we find oscillations in the wake 
structure behind a fast moving color charge akin to Cerenkov like 
radiation and Mach cone. We further show that the wake potential for 
incoming flow is much like a Coulomb potential, while the 
potential in the outward flow flips its sign due to dynamical screening. 
%When static screening is absent a similar effect was found in a 
%system of gravitating masses~\cite{Bashkirov98} and also in a 
%complex dusty plasma~\cite{Bashkirov04,Tsytovich03} within the weak coupling 
%limit~\cite{Ichimaru82},
%which leads to a long range correlation between plasma particles. Depending 
%upon the nature of the potential (strength of the interaction) the system could be in liquid or 
%crystalline 
%states~\cite{Bashkirov04} and such states were also observed earlier
%in various laboratories for dusty plasmas~\cite{Labplasma}. 
In addition a potential perpendicular to the motion of the parton is
also calculated which indicates a transverse flow in the QGP.
We further compute a color dipole potential considering two moving charges
and discuss the consequences of possible new bound states, {\it viz.}, 
colored bound states, and $J/\psi$ suppression in the QGP. 
%Further, in the same spirit, 
%we obtain the structure and pair correlation functions to understand
%the structure of the plasma.

\section{Response of the Quark-Gluon Plasma}
\label{sec_res_QGP}
An appropriate description~\cite{Ichimaru73,Landau81} of various plasma 
properties
can be obtained if one knows how a plasma responds to an external disturbance.
In order to establish a response relationship in a plasma, one usually 
considers the plasma response to the external electric field which induces a
current density. If the system is stable against such a disturbance whose
strength is weak, then the induced current may appropriately be expressed
by that part of the response which is linear in the externally 
disturbing field. The {\it linear response} of a plasma to an external 
electromagnetic field has extensively been studied~\cite{Ichimaru73,Landau81}
in plasma physics in which the external current is related to the 
total electric field by
\begin{equation}
\vec{J}^a_{\rm{ind}}(\omega,\vec {\mathbf k})= -\frac{i\omega}{4\pi}
\left [ \epsilon (\omega,\vec {\mathbf k})- \mathbb{\openone} \right ] 
\vec{E}^a_{\rm{tot}} (\omega,\vec {\mathbf  k}) , \label{ejrel} 
\end{equation}
where $\epsilon (\omega,\vec {\mathbf k})$ is the dielectric tensor,
describing the linear (chromo)electromagnetic properties of the medium
and $\mathbb{\openone}$ is the identity operator.
Since we are interested in a system of a relativistic
color charge moving through a QCD plasma, we apply the linear response
theory for the purpose in the same spirit as in~\cite{Ruppert} by simply 
assigning a color index, $a=1\cdots 8$, to the relevant quantities here 
and also in the subsequent analysis to take into account the
quantum and non-Abelian effects. Non-Abelian effects (beyond the 
color factors, e.g., in the Debye mass) will be important at realistic
temperatures. Unfortunately, they cannot be treated by the method used
by Ruppert and M\"uller~\cite{Ruppert} and are therefore beyond the scope 
of this work as we closely follow their work in our analysis.   
 
\subsection{Linear Response in a (Color) Plasma} 
\label{sec_linqgp}

In this subsection we briefly outline the theoretical 
description~\cite{Ichimaru73,Ruppert} based on the linear response theory 
for the QGP as a finite, continuous, homogeneous and isotropic dielectric medium,
which can be expressed by a dielectric tensor, 
$\epsilon(\omega,\vec {\mathbf k})$, depending on the direction only
through the momentum vector, $\vec {\mathbf  k}$. One can also construct
another set of tensors of rank two from the momentum vector, 
 $\vec {\mathbf  k}$ which are the longitudinal 
projection tensor, ${\cal P}^L$ and the transverse projection tensor, 
${\cal P}^T$, respectively, given as 
\begin{eqnarray}
{\cal P}^L_{ij} &=& \frac{k_ik_j}{k^2} \ \ , \nonumber \\
{\cal P}^T_{ij} &=& \delta_{ij}- \frac{k_ik_j}{k^2} \ \ , \label{proj}
\end{eqnarray}
with the general properties of the projection operators:  
$({\cal P}^L)^2 ={\cal P}^L$, $({\cal P}^T)^2 ={\cal P}^T$,
${{\cal P}^L}\cdot {{\cal P}^T} =0$ and
$({\cal P}^L)^2 + ({\cal P}^T)^2 =1$.
Now the dielectric tensor can be written  as a linear combination of 
these two mutually independent components as
\begin{eqnarray}
\epsilon(\omega,\vec {\mathbf k}) &=& {\cal P}^L\epsilon_L(\omega,k)
\ + \ {\cal P}^T\epsilon_T(\omega,k) \, \, , \nonumber \\
\epsilon_{ij} &=& {\cal P}^L_{ij}\epsilon_L (\omega,k)\ 
+ \ {\cal P}^T_{ij}\epsilon_T(\omega,k) \ , \label{diele} 
\end{eqnarray}
where the longitudinal, $\epsilon_L$ and the transverse, $\epsilon_T$, 
dielectric functions are given by
\begin{eqnarray}
\epsilon_L(\omega,k) &=& \frac{\epsilon_{ij}k_ik_j}{k^2} \, \, , \nonumber\\
\epsilon_T(\omega,k) &=& \frac{1}{2}\left [{\rm {Tr}} 
\epsilon(\omega,\vec{\mathbf k})
-\epsilon_L (\omega,k) \right ] \, \, . \label{diele1}
\end{eqnarray} 
Now, we shall use the dielectric tensor in (\ref{diele}) for a
set of macroscopic equations, {\it viz.}, Maxwell and continuity equations,
in momentum space to obtain a relation~\cite{Ichimaru73} between the
total chromoelectric field and the external current as
\begin{eqnarray}
\left [\epsilon_L {\cal P}^L +\left ( \epsilon_T -\frac{k^2}{\omega^2}
\right ){\cal P}^T \right ] \vec{E}^a_{\rm {tot}}(\omega,\vec{\mathbf k})= 
\frac{4\pi} {i\omega} \vec{J}^a_{\rm {ext}}(\omega,\vec{\mathbf k}) 
\, \, . \label{extj}
\end{eqnarray}
If there is no external disturbance to the plasma, $\vec{J}^a_{\rm{ext}}=0$, 
(\ref{extj}) reduces to
\begin{eqnarray}
\left [\epsilon_L {\cal P}^L +\left ( \epsilon_T -\frac{k^2}{\omega^2}
\right ){\cal P}^T \right ] \vec{E}^a_{\rm {tot}}(\omega,\vec{\mathbf k})=0 
\, \, , \label{extj1}
\end{eqnarray}
which has nontrivial solutions representing dispersion relations in a linear
medium, if the determinant vanishes, i.e.,
\begin{eqnarray}
\det \left | \epsilon_L {\cal P}^L +\left ( \epsilon_T -
\frac{k^2}{\omega^2} \right ){\cal P}^T  \right |=0 
\, \, , \label{detm}
\end{eqnarray}
leading to the following equations defining the longitudinal and transverse modes
\begin{eqnarray}
\epsilon_L(\omega,k)=0 \, \, , \nonumber \\
\epsilon_T(\omega,k)= \frac{k^2}{\omega^2} \, \, . \label{dispg}
\end{eqnarray}
So, the dielectric tensor contains essentially all the information 
of the chromoelectromagnetic properties of the plasma. However,
one can also study a density response~\cite{Ichimaru73} to an 
external test-charge field through a density fluctuation which could induce
a color charge density as
\begin{equation}
\rho^a_{\rm{ind}}(\omega,k) =\left (\frac{1}
{\varepsilon(\omega,\vec{\mathbf k})}
-1\right ) \rho^a_{\rm {ext}}(\omega,k)  \, \, , \label{indrho}
\end{equation} 
where $\rho^a_{\rm{ext}}$ is the external charge density and
$\varepsilon(\omega,\vec{\mathbf k})$ is an another dielectric response 
function that can be obtained as
\begin{equation}
\varepsilon(\omega,\vec{\mathbf k}) = \frac{\vec{\mathbf k}\cdot\varepsilon
(\omega,\vec{\mathbf k}) \cdot \vec{\mathbf k}} {k^2} \, \, .
\label{altdiel}
\end{equation} 
This is analogous to the longitudinal dielectric function in (\ref{diele1}) 
with the difference that it is independent of the direction of $\vec{\mathbf k}$ 
because of isotropy. For a homogeneous and isotropic system, one can write
\begin{equation}
\epsilon_L(\omega, k)=\varepsilon(\omega,\vec{\mathbf k}) \,
\, , \label{equvdiel}
\end{equation} 
which can lead to  a longitudinal dispersion relation for density
fluctuations indicating that the space charge field could be spontaneously
excited without any external disturbance. 

Combining (\ref{equvdiel}) with (\ref{indrho})
the induced color charge density by  the external charge distribution becomes
\begin{equation}
\rho^a_{\rm{ind}}(\omega,k) =\left (\frac{1}{\epsilon_L(\omega, k)}-1\right )
\rho^a_{\rm {ext}}(\omega,k)  \, \, , \label{indrho1}
\end{equation} 
and the total color charge density is given as
\begin{equation}
\rho^a_{\rm {tot}}(\omega,k)= \rho^a_{\rm {ind}}(\omega,k) \ + \
\rho^a_{\rm {ext}}(\omega,k)= \frac{\rho^a_{\rm {ext}}(\omega,k)}
{\epsilon_L(\omega,k)} \, \, , \label{totchg}
\end{equation} 
in which the dielectric function measures the screening of an 
external charge due to the induced space charge in the plasma.
 
The screening potential in momentum space can be obtained 
from Poisson equation in  Coulomb gauge~\cite{Ichimaru73,Landau81} as
\begin{equation}
\Phi^a(\omega,k) = 4\pi \frac{\rho^a_{\rm {ext}}(\omega,k)} 
{k^2 \epsilon_L(\omega,k)} \, \, .  \label{pot} 
\end{equation}

Combining (\ref{ejrel}), (\ref{diele}) and (\ref{extj}) the induced current
density can be related to the external current density as
\begin{equation}
\vec {J}^a_{\rm{ind}} (\omega,\vec{\mathbf k}) = \left [ \left ( 
\frac{1}{\epsilon_L}-1 \right ){\cal P}^L +\frac{1-\epsilon_T}
{\epsilon_T-k^2/\omega^2} {\cal P}^T \right ] 
\vec{J}^a_{\rm{ext}}(\omega,\vec{\mathbf k}) \, \, .\label{relinex}
\end{equation}

\subsection{A Fast Color Charge in a Plasma}
\label{fast_parton}

Since we are interested in studying the wake behavior of the plasma reacting to
a charged particle, $Q^a$ moving with a constant velocity $\vec {\mathbf v}$, 
we
specify the external current and charge density, respectively, as
\begin{eqnarray}
\vec{J}^a_{\rm{ext}}&=& Q^a \vec {\mathbf v} \delta (\vec{\mathbf x}-
\vec{\mathbf v}t ) \stackrel{\rm{FT}}{=} 2\pi Q^a \vec{\mathbf v} 
\delta(\omega - \vec{\mathbf k}\cdot \vec{\mathbf v}) \, \, \, , \nonumber \\ 
\rho^a_{\rm{ext}}&=& Q^a  \delta (\vec{\mathbf x}-
\vec{\mathbf v}t ) \stackrel{\rm{FT}}{=} 2\pi Q^a  
\delta(\omega - \vec{\mathbf k}\cdot \vec{\mathbf v}) \, , \label{initchgj} 
\end{eqnarray} 
where ${\rm{FT}}$ stands for Fourier transformation and $v=|\vec{\mathbf v}|$.
The finite motion of
the charged particle deforms the screening charge cloud and also suffers a 
retarding force causing energy-loss. The soft contribution to the
collisional differential energy-loss by the induced chromoelectric field 
is defined as~\cite{Ichimaru73,Thoma91}
\begin{equation}
\frac{dE}{dx} = Q^a \frac{\vec{\mathbf v}}{v} {\rm {Re}}\vec{E}^a_{\rm{ind}}
(\vec{\mathbf x}=\vec{\mathbf v}t,t) \, \, , \label{diffel}
\end{equation}
The induced electric field can be obtained from the inverse of (\ref{diele1})
which is given as
\begin{equation}
\vec{E}^a_{\rm{ind}} = \left [ \left ( \frac{1}{\epsilon_L}-1\right ){\cal P}^L 
+\left(\frac{1}{\epsilon_T-k^2/\omega^2}-\frac{1}{1-k^2/\omega^2}  \right)  
{\cal P}^T \right ]\frac{4\pi}{i\omega} \vec{J}^a_{\rm{ext}} 
\, \, . \label{eind}
\end{equation}
Combining (\ref{diffel}) and (\ref{eind}) the soft contribution to the
differential collisional 
energy-loss in QGP can be obtained~\cite{Thoma91} as
\begin{equation}
\frac{dE}{dx} = - \frac{C\alpha_s}{2\pi^2v} \int d^3k \frac{\omega}{k^2}
\left [ {\rm{Im}}\epsilon^{-1}_L + \left (v^2k^2-\omega^2\right )
{\rm{Im}}\left (\omega^2\epsilon^{-1}_T -k^2 \right )^{-1}\right ] \, \,
, \label{cel}
\end{equation}
where $\omega=\vec{\mathbf k}\cdot \vec{\mathbf v}$ originates from 
(\ref{initchgj}), $C$ is the quadratic Casimir invariant, and 
$\alpha_s=g^2/4\pi$ is the strong coupling constant.

\subsection{Response Functions in the High Temperature Limit}
\label{resf_ht}

The dielectric functions, both longitudinal and transverse, in (\ref{diele1})
are related~\cite{Kapusta,Lebellac,Thoma91,Weldon82,Elze89,Mrow90} 
to the self-energies of the gauge
boson, i.e., gluon, in the medium as
\begin{eqnarray}
\epsilon_L(\omega,k) &=& 1-\frac{\Pi_L(\omega,k)}{K^2} \, \, , \nonumber \\
\epsilon_T(\omega,k) &=& 1-\frac{\Pi_T(\omega,k)}{\omega^2} \, \, ,
\label{polari}
\end{eqnarray}
where $K=(\omega,k)$ with $k=|\vec{\mathbf k}|$. $\Pi_L$ and $\Pi_T$ are, 
respectively, the longitudinal and the transverse self energies of the gluon,
which are to be calculated in the high temperature limit.

The transversality of the polarization tensor, $K^\mu\Pi_{\mu\nu}=0$, implies 
that only two components
are independent, and the general solution can be written as
\begin{equation}
\Pi_{\mu\nu}(\omega,\vec{\mathbf k})= {\cal A}_{\mu\nu}\Pi_T(\omega,k)
+{\cal B}_{\mu\nu}\Pi_L(\omega,k) \, \, , \label{genpi}
\end{equation} 
where the transverse and longitudinal projection 
tensors~\cite{Weldon82,Das} are
\begin{eqnarray}
{\cal A}_{\mu\nu} &=& {\tilde{\eta}}_{\mu\nu}- 
\frac{\tilde{K}_\mu \tilde{K}_\nu} {\tilde{K}^2} \, \,  , \nonumber \\
{\cal B}_{\mu\nu} &=& -\frac{1}{{{K}}^2\ k^2}\left [ 
\left( k^2u_\mu +\omega {\tilde K}_\mu \right)
\left( k^2u_\nu +\omega {\tilde K}_\nu \right) \right ] \, \, , \label{ltproj}
\end{eqnarray}
in which $u_\mu$ is the four velocity of the fluid with $u^\mu u_\mu=1$. The tensor
$\tilde{\eta}_{\mu\nu}$ and the vector $\tilde{K}$ which are orthogonal 
to $u_\mu$ are defined as
\begin{eqnarray}
\tilde{\eta}_{\mu\nu} &\equiv& {\eta}_{\mu\nu} -u_\mu u_\nu \, \, ,
\nonumber \\
\tilde{K}_\mu &\equiv& K_\mu -\omega u_\mu \, \, , \label{orthotv}
\end{eqnarray}
with the Minkowski metric tensor $\eta_{\mu \nu}$.
Using the properties of the projection operators, one can obtain the
scalar functions in (\ref{genpi}) as
\begin{eqnarray}
\Pi_L(\omega,k) &=&  - \frac{K^2}{k^2}\Pi_{00} (\omega,k) \, \, , \nonumber \\ 
\Pi_T(\omega,k) &=&  \frac{1}{2}\left (\delta_{ij}-\frac{k_ik_j}{k^2}
\right ) \Pi_{ij} (K) = \frac{1}{2} \left [ \Pi^\mu_\mu(\omega,k)
-\Pi_L(\omega,k) \right ] \, \, . \label{selfg} 
\end{eqnarray}
For computing the self energies one needs the in-medium gluon propagator 
which can be obtained from the Dyson-Schwinger equation
\begin{equation}
(\Delta_{\mu\nu}(K))^{-1} = (\Delta^0_{\mu\nu}(K))^{-1} + \Pi_{\mu\nu}(K)
\, \, , \label{dyson}
\end{equation}
where $\Delta^0_{\mu\nu}(K)$ is the gluon propagator in vacuum.
Combining (\ref{genpi}), (\ref{ltproj}) and (\ref{dyson}), the full 
in-medium gluon propagator in covariant 
gauge~\footnote{This particular gauge is chosen for convenience as we will
see later in subsec.~\ref{dipole_pot}.} 
is obtained as
\begin{equation}
\Delta_{\mu\nu}(K) = - \frac{{\cal A}_{\mu\nu}}{K^2-\Pi_T}
 - \frac{{\cal B}_{\mu\nu}}{K^2-\Pi_L}
 +(\xi -1) \frac{K_\mu K_\nu}{K^4} \, \, , \label{fullprop}
\end{equation} 
where $\xi$ is the gauge parameter.

The one loop gluon self energy in the high temperature limit has been 
obtained in Refs.~\cite{Klim82,Weldon82} which was found to be 
equivalent~\cite{Lebellac,Thoma00} to the HTL approximation~\cite{Braaten90} 
that relies on the restriction to hard loop momenta and is given as
\begin{eqnarray}
\Pi_L(\omega,k) &=& -m^2_D \frac{K^2}{k^2} \left [ 1- \frac{\omega}{2k} \ln 
\frac{\omega+k}{\omega-k} \right ] \, \ , \nonumber \\
\Pi_T(\omega,k) &=& \frac{m^2_D}{2}\frac{\omega^2}{k^2}
 \left [ 1- \left (1-\frac{\omega^2}{k^2} \right )
\frac{\omega}{2k} \ln 
\frac{\omega+k}{\omega-k} \right ] \, \, , \label{selfg1}
\end{eqnarray}
where the Debye screening mass follows from
\begin{equation}
m^2_D =\Pi_L(\omega=0,k)= \Pi_{00}(\omega=0,k) 
= g^2T^2\left(1+\frac{N_f}{6}\right ) \, \, ,
\label{sceenm}
\end{equation} 
in which $N_f$ is the number of light quark flavors in the QGP. Though
the gluon propagator is a gauge dependent quantity, the gluon self energies
in (\ref{selfg1}) are gauge independent  only for the leading
term of the high temperature expansion~\cite{Heinz87}. The dielectric functions 
are therefore also gauge invariant and can be written combining (\ref{polari})
 and (\ref{selfg1}) as
\begin{eqnarray}
\epsilon_L(\omega,k)&=& 1+\frac{m^2_D}{k^2} \left [ 1- \frac{\omega}{2k} 
\left ( \ln \left |\frac{\omega+k}{\omega-k}\right | 
-i\pi \Theta(k^2-\omega^2) \right ) \right ] 
\, \ , \nonumber \\
\epsilon_T(\omega,k) &=& 1- \frac{m^2_D}{2\omega^2}
\left [ \frac{\omega^2}{k^2}
 - \left (1-\frac{\omega^2}{k^2} \right )
\frac{\omega^3}{2k^3} \left ( \ln \left | 
\frac{\omega+k}{\omega-k}\right | -i\pi \Theta(k^2-\omega^2) \right ) \right ] 
\, \, . \label{htldiel}
\end{eqnarray}

\begin{figure}
\begin{center}
\includegraphics[width=0.6\linewidth]{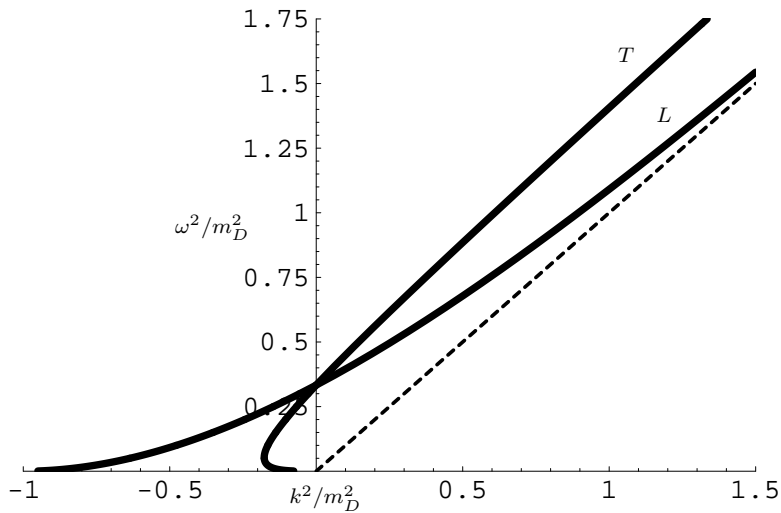}
\setlength{\unitlength}{1cm}
\begin{picture}(6,0)
\put(-6.5,0.1){\makebox(0,0){\footnotesize $k^2/m_D^2$}}
\put(-8,3.7){\makebox(0,0){\footnotesize $\omega^2/m_D^2$}}
\put(-2.0,5.2){\makebox(0,0){\footnotesize $L$}}
\put(-2.5,6){\makebox(0,0){\footnotesize $T$}}
\end{picture}
\caption{Dispersion relations for the longitudinal ($L$) 
and the transverse ($T$) part  of gluons in the QGP (quadratic scale). 
The dashed line indicates the light cone, $\omega=k$ .}
\label{fig:gluonmodes1}
\end{center}
\end{figure}

The dispersion of longitudinal and transverse gluons, determined
by the respective real parts of the dielectric functions in (\ref{dispg}) 
(also can be obtained from the poles of the gluon propagator in 
(\ref{fullprop})), are shown in Fig.~\ref{fig:gluonmodes1}. The following
important features of the plasma are transparent from 
Fig.~\ref{fig:gluonmodes1}: 
{\bf (i)} There are propagating modes above a common plasma frequency 
$\omega > \omega_{\rm{pl}} =\sqrt{m^2_D/3}$, which are always above the 
light cone ($\omega > k$), show collective behavior, and are called normal 
modes.  The longitudinal mode (plasmon) propagates with an average phase
velocity $\sqrt{3/5}c$ whereas the transverse mode with 
$\sqrt{6/5}c$ for moderate $k$~\cite{Weldon82}.
{\bf(ii)} In the static limit $\omega=0$, the self energies $\Pi_L(0,k)=m^2_D$
and $\Pi_T(0,k)=0$, respectively. This means that the longitudinal
branch indicates the static screening in the electric sector
with an inverse screening length, $m_D$, given in (\ref{sceenm}), 
whereas the solution of the transverse branch at $k=0$, implies the absence
of magnetostatic screening. 
{\bf (iii)} For $0< \omega < \omega_{\rm{pl}}$, $k$ becomes
imaginary~\cite{Weldon82} and there are no stable normal modes as in (i), 
however, there is a collective behavior which corresponds to dynamical 
screening both in electric and magnetic sector as long as $\omega > 0$. 
{\bf (iv)} For $\omega < k$, the imaginary parts of the dielectric functions 
corresponds to Landau damping causing energy dissipation in the plasma
through elastic collisions.

When a color charge  moves with a velocity, $\vec{\mathbf v}$, the dispersion
relation disappears from above the light cone and owing to the interaction 
of the moving test charge with the medium, it is possible 
to define another mode as
\begin{equation}
\omega=\vec{\mathbf k} {\mathbf \cdot} \vec{\mathbf v} \, \, \label{cheren}
\end{equation}
which has the support from (\ref{initchgj}) due to the $\delta$-function 
restriction.  The collective modes in 
(\ref{cheren}) are in the {\it space like region} in which ${\rm Re}(\omega) <k$ 
for $0<v<1$ in contrast to the normal collective modes, where 
${\rm Re}(\omega) >k$ always holds as discussed above in case (i). 
The dielectric response functions in (\ref{htldiel}) provide a direct
measure of the extent of which an external test charge is screened by
the induced space charge in the plasma. The screening in this case is 
{\it dynamical} in the sense that it depends on the frequency as well as
the wave vector similar to  the case (iii) with the difference that
the wave vector is damped~\cite{Weldon82} there. 

Equation (\ref{cheren}) is also known as {\it 
Cerenkov condition}~\cite{Ichimaru73} for wave emission by a moving 
parton in a plasma. Since the average phase velocity of the
transverse modes $\sqrt{6/5}c$ is greater than the speed of light for
the isotropic plasma, the Cerenkov condition is not satisfied.
According to (\ref{cheren}) there could be two important effects 
due to the particle interaction with the plasmon wave, which is propagating with 
an average phase velocity $\sqrt{3/5}c=0.77c$:

1) The modes which are moving with a speed less than the average speed of the
plasmon mode ($0.77c$) can be excited but they accelerate 
the slower moving charge particles and decelerate those moving faster 
than the modes. In the former case energy of the excited modes is 
absorbed whereas in latter case energy is transfered to those modes by
the charges. 
Such modes are excited by a charged particle which moves less than 
$0.77c$ but are severely Landau damped owing to absorption and emission
resulting in a wake in the induced charge density as well as in the potential.

2) The modes which are moving with a speed greater than 
$0.77c$ can be excited and they may not be damped,
which are different from the existent plasma waves as discussed in case 1). 
When a charged particle moves faster than $0.77c$, it can excite
undamped modes which could generate Cerenkov like radiation
and Mach shock waves leading to oscillations in the induced charge density and
in the corresponding potential. We note that Mach waves are generated in the 
plasma in general if the particle moves faster than the speed of sound, 
$c_s$ in the plasma and the collective modes shift to the space like 
region. The Mach opening angle is given by
\begin{equation}
\phi_M= \arccos\left ( \frac{c_s}{v}\right ) \, \, . \label{mangle}
\end{equation}
%where $c_s$ is the speed of of sound in the plasma.

In the following we would like to investigate
those effects of the dynamical screening due to the fast moving
charged particle on the wake of a color plasma in the high temperature limit. 

\subsection{The Wake in the Induced Charge Density}
\label{chg_ind}

Combining (\ref{indrho1}) and (\ref{initchgj}),
the induced charge density in configuration space reads as~\cite{Mustafa05}
\begin{equation}
\rho^a_{\rm{ind}}(t,{\vec{\mathbf r}})= 2\pi Q^a \
\int \frac{d^3k}{(2\pi)^3} \int \frac{d\omega}{2\pi} 
\exp\left [ i\left ({\vec{\mathbf k}} \cdot {\vec {\mathbf r}} - \omega t
\right )\right ] 
\left [\frac{1}{\epsilon_L(\omega,k)}-1 \right ]\ \delta(\omega -{\vec
{\mathbf k}} \cdot {\vec {\mathbf v}}) \, \, 
. \label{indrho2}
\end{equation}

In cylindrical coordinates ${\vec {\mathbf k}}= (\kappa \cos\phi, 
\kappa \sin \phi, k_z)$, ${\vec{\mathbf r}}= (\rho,0,z)$ and assuming
${\vec{\mathbf v}}=(0,0,v)$, the induced charge density in (\ref{indrho2})
becomes
\begin{eqnarray}
\rho^a_{\rm{ind}}(t,\rho,z) &=& \frac{Q^a}{(2\pi)^2v} \int \kappa \ d\kappa 
\ J_0 (\kappa \rho)\ \int_{-\infty}^{+\infty}\ d\omega \, 
\exp \left [ i\omega \left (\frac{z}{v}-t \right )\right ]\,
\left [\frac{1}{\epsilon_L(\omega,k)}-1 \right ] \, \, , \label{indrho3}
\end{eqnarray}
where $J_0$ is the Bessel function and 
$k=\sqrt{\kappa^2+\omega^2/v^2}$.

Using the symmetry properties of $\epsilon_L(\omega,k)$, {\it viz.},
\begin{eqnarray}
{\rm {Re}}\ \epsilon_L(-\omega,k)&=& {\rm {Re}}\ \epsilon_L(\omega,k) \, \, ,
\nonumber \\
{\rm {Im}}\ \epsilon_L(-\omega,k)&=& -{\rm {Im}}\ \epsilon_L(\omega,k)  ,
\label{symdiel}
\end{eqnarray}
the induced charge density in (\ref{indrho3}) reads as
\begin{eqnarray}
\rho^a_{\rm{ind}}(t,\rho,z) &=& \frac{Q^a}{(2\pi)^2v} \int \kappa \ d\kappa 
\ J_0 (\kappa \rho)\ \int_{0}^{\infty}\ d\omega \, 
\left [\cos \left ( \omega \left (\frac{z}{v}-t\right )\right ) 
\left ( \frac{{\rm{Re}}\ \epsilon_L}{\Delta} - 1 \right ) 
+ \sin \left ( \omega \left (\frac{z}{v}-t\right ) \right )
\frac{{\rm{Im}}\ \epsilon_L}{\Delta}  
\right ] \, \, , \label{indrho4}
\end{eqnarray}
where $\Delta=[{\rm{Re}}\ \epsilon_L]^2+ [{\rm{Im}}\ \epsilon_L]^2$

Now we would like to point out that, if the external charge is at rest, 
$\vec{\mathbf v}\rightarrow 0$, relative
to the background, the induced charge cloud remains spherically symmetric.
In the limit $\vec{\mathbf v}\rightarrow 0$, (\ref{indrho4}) reduces to
\begin{eqnarray}
\rho^a_{\rm{ind}}(\rho,z)&=& -\frac{m^2_D Q^a}{4\pi} \int_0^{\infty} 
d\kappa \frac{J_0(\kappa \rho)}{\sqrt{\kappa^2+m_D^2}} 
\exp \left [ -z {\sqrt{\kappa^2+m_D^2}} \right ] \, \, , \nonumber\\
&=& - \ \frac{Q^a}{4\pi} m^2_D \frac{\exp[-m_D r]}{r} \, . \label{ind_yukawa}
\end{eqnarray}
This can be understood as follows: the background particles are moving
isotropically on the average around the test charge. Introduction of a
test charge which is at rest merely introduces a local fluctuation of
the number density in its vicinity but does not spoil the spherical 
symmetry of the system. As a result the response of the medium 
modifies the induced charge density into a static Yukawa type still 
reflecting this symmetry.

The situation, however, changes if the test charge is in motion relative
to the heat bath. The motion of the particle fixes the direction in space
and spherical symmetry of the problem reduces to axial symmetry. This
implies the loss of spherical symmetry of the Debye screening cloud
around the moving test charge resulting in a wake in the induced charge
due to dynamical screening as given in (\ref{indrho4}). 

\begin{figure}
%\begin{minipage}[h]{0.48\textwidth}
\begin{minipage}[h]{0.48\textwidth}
\centering{\includegraphics[height=0.8\textwidth,width=1.2\textwidth]
{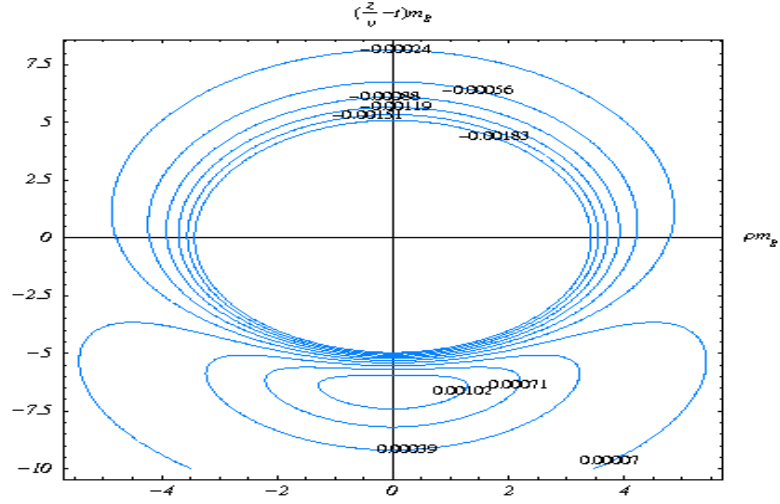}}
\end{minipage}
%\vspace{0.8in}
\caption{The plot shows the equicharge lines in the induced charge density 
for $v=0.55c$ with a fixed color charge $Q^a$ in the high temperature plasma. }
\label{fig:indchg_v55}
\end{figure}

In Fig.~\ref{fig:indchg_v55} we display the equicharge lines in the induced
charge density for $v=0.55c$ in cylindrical coordinates which are scaled 
with $m_g=\sqrt{m_D^2/3}$. (The induced charge density is proportional to 
$m_g^3$.) The induced charge density is found to carry a screening charge cloud
with a sign flip along the direction of the moving charge. As discussed in 
the preceding subsection~\ref{resf_ht} there will be an excitation of modes
during such a wake but they will be severely Landau damped through 
induced emission and absorption of the plasmons by scattering off single
particles. As a result the particle suffers an energy loss through elastic
collisions which can be described by (\ref{cel}) and has been studied
extensively in the literature~\cite{Thoma91}. When a charged particle
moves with a velocity of less than $0.77c$, $\gamma v\leq 2.5$, the soft 
part of the collisional energy loss, caused by the induced chromoelectric
field, turns out to be important~\cite{Munshi,Jane,Moore05,Munshi05}. 

\begin{figure}
%\begin{minipage}[h]{0.48\textwidth}
\begin{minipage}[h]{0.48\textwidth}
\centering{\includegraphics[height=0.8\textwidth,width=0.8\textwidth]
{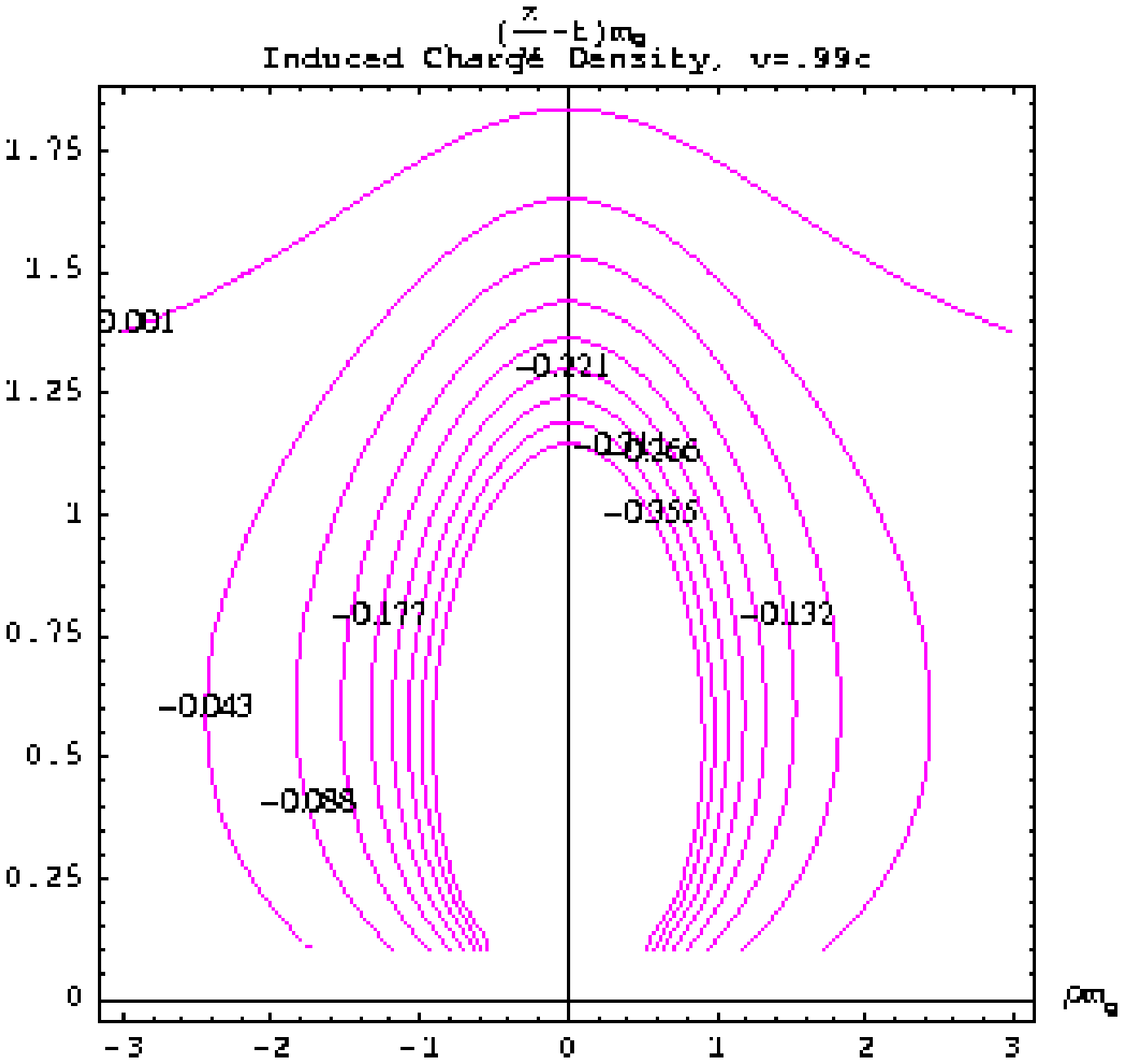}}
\end{minipage}
\begin{minipage}[h]{0.48\textwidth}
\centering{\includegraphics[height=0.8\textwidth,width=0.8\textwidth]
{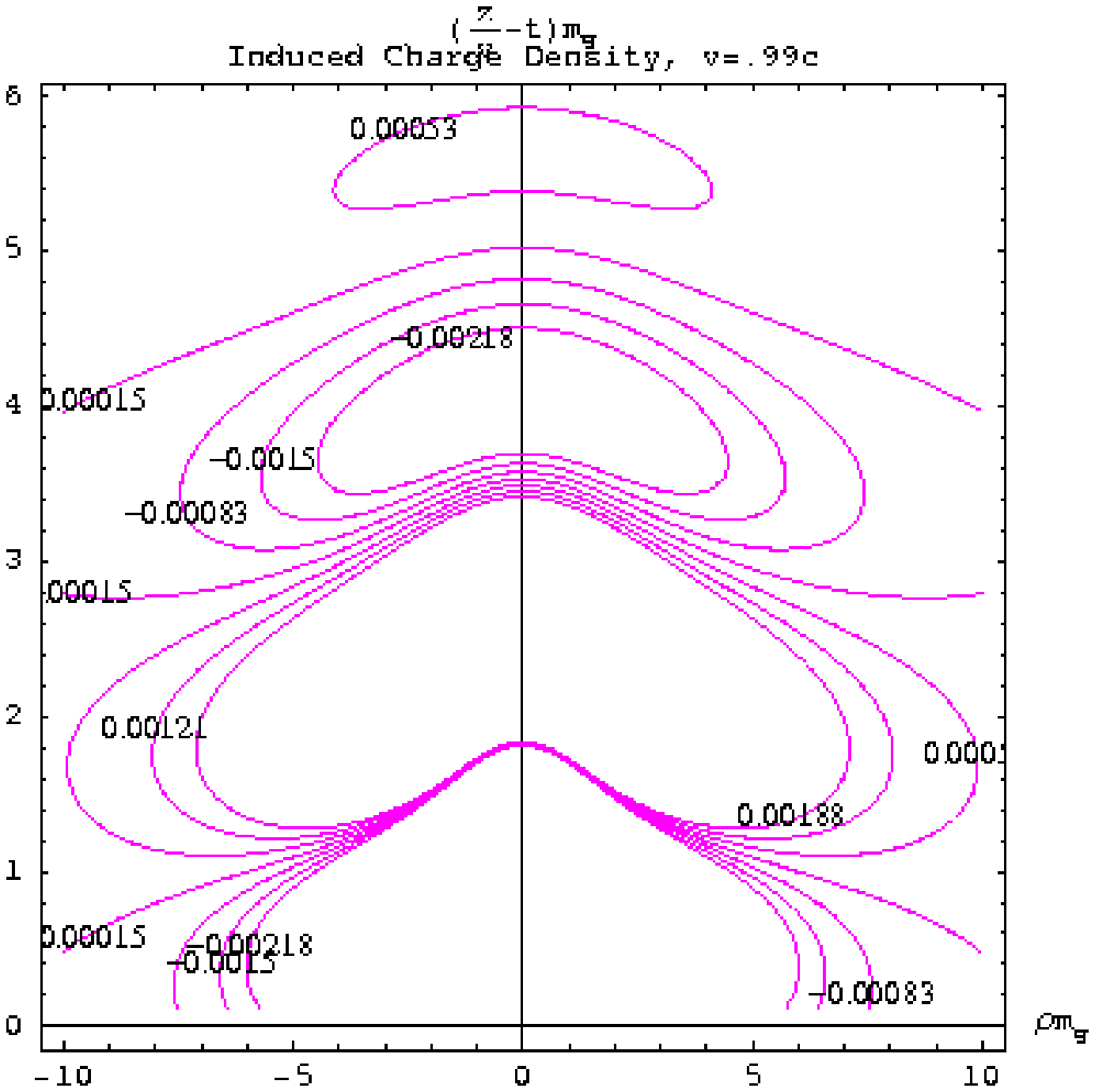}}
\end{minipage}
%\vspace{0.8in}
\caption{Left panel: The plot reproduces the equicharge lines in the 
induced charge density for $v=0.99c$ with a fixed color charge 
$Q^a$ in the high temperature plasma as obtained in Ref.~\cite{Ruppert}. 
Right panel: Same as left panel but plotted with more number of equicharge
lines with enlarged axes to show the oscillations in the same density wake.}
\label{fig:indchg_v99}
\end{figure}

Next we consider the case when a charged particle moves faster than
the average speed of the plasmon. For this purpose we choose $v=0.99c$ and
the result is shown in the left panel of Fig.~\ref{fig:indchg_v99}
which reproduces the corresponding results of Ref.~\cite{Ruppert}. 
The wake in the induced charge density apparently 
does not show any sign flip as in Fig.~\ref{fig:indchg_v55} and it 
shows  merely a screening cloud around a test charge similar to 
that given in (\ref{ind_yukawa}). However, this is not the case as 
the particular analysis in ~\cite{Ruppert} was not 
complete and some important features were obviously missing in the 
wake of the medium. 

The complete analysis is shown in the right panel of 
Fig.~\ref{fig:indchg_v99} when the same data corresponding to the left 
panel are plotted with enlarged axes with a larger number of equicharge 
lines in the induced charge density. As expected the wake of the medium
is clearly visible in the induced charge density which follows the  
charged particle traveling with a velocity, $v=0.99c$. Moreover, the
screening cloud following the charge particle is found to flip its sign
alternately indicating enhanced and depleted charge density. 
As discussed in the preceding subsection~\ref{resf_ht} 
there will be an excitation in the plasmon modes during such a wake which 
could emit a Cerenkov like radiation with a Mach cone structure trailing 
the moving parton. The radiation is like a spontaneous emission 
of plasmon waves and cannot be calculated through the expression given in 
(\ref{cel}) and requires a special treatment. It is also important to note
that for a parton velocity $v=0.99c$, $\gamma v \geq 4$, corresponding to 
the ultra-relativistic limit~\cite{Moore05,Munshi05} for which the hard contribution
of the collisional energy loss due to elastic scattering and the radiative 
energy loss are expected to be dominant.

When a parton moves supersonically, it excites waves in a colored
plasma. The propagation of sound waves could be estimated from the emission
pattern of secondary particles traveling at an angle with respect to the
jet axis through the relation~\cite{Stocker,Ruppert}
\begin{equation}
\Delta \phi = \pi \pm \phi_M \,= \pi \pm \arccos\left ( \frac{c_s}{v}\right ) 
\, , \label{opening}
\end{equation}
where $\Delta \phi=\pi$ corresponds to the location of a maximum distribution 
of secondary hadrons from the away side jet in $p+p$ collisions, where no 
medium effects are present. However, due to medium effects the azimuthal 
angle for secondaries is expected to be peaked differently from $\pi$  
and could be related to the Mach opening angle, $\phi_M$. 

There are several programs in STAR and PHENIX 
experiments at RHIC where various approaches are being followed in order to 
find Mach cones. The basic approach is to use correlations between high $p_\bot$ 
secondary particles in pseudorapidity and azimuthal space. Some of the 
background correlation effects which might mimic Mach cones are being 
studied in details. Both the experiments
are looking at 3-particle correlations which will eliminate the trivial 
backgrounds. Now already some interesting results have been reported by the 
PHENIX collaboration~\cite{Azimuthal} where the data suggest that the peak in 
the secondary particle correlation provides compelling evidence for a
strong modification of the away-side jet. This could indicate two possible
scenarios: (i) a Cerenkov jet in which the leading and away-side jet axes are
aligned but fragmentation is confined to a very thin hollow cone or (ii) a
directed jet in which the away-side jet axes are misaligned. A more quantitative
investigation is necessary to distinguish between a $`$Cerenkov jet' that 
leads to a Mach cone, $\phi_M$ and a $`$directed jet' which is misaligned. 
Following (\ref{opening}) the azimuthal angle can be estimated here as 
$\Delta \phi\approx \pi\pm 1.01$ 
using $c_s=0.48$ within the HTL approximation. 
Once a characteristic correlation structure in secondary hadrons leading to
a Mach cone, $\phi_M \sim 1.01$ is 
confirmed experimentally, it would provide an estimation of the angular 
structure of the energy loss and also the speed of sound in 
the QGP. 

In Ref.\cite{Flow} it was argued that the wake of an energetic parton
can create an observable flow in the QGP. However, it was also argued
that this is only the case if the energy loss of the parton is very large
(about 12 Gev/fm). Even combining the collisional and radiative contributions,
we do not expect such an high energy loss in relativistic heavy-ion collisions.

\subsection{The Wake in the Screening Potential}
\label{wake_pot}

Combining (\ref{pot}) and (\ref{initchgj})  the screening potential
in configuration space due to the motion of a color charge can be
written as \cite{Mustafa05}
\begin{equation}
\Phi^a\left(\vec{\mathbf r},\vec{\mathbf v},t\right) = \frac{Q^a}{2\pi^2}\int
d^3k \frac{e^{i\vec{\mathbf k}\cdot\left(\vec{\mathbf r}-\vec{\mathbf v}
t\right)}}{k^2 \epsilon_L
\left(\omega=\vec{\mathbf v}\cdot\vec{\mathbf k},k\right)}\,.\label{pot1}
\end{equation}

In cylindrical coordinates, the screening potential in (\ref{pot1}) becomes
\begin{eqnarray}
\Phi^a\left(\rho,z,t\right) &=& \frac{2Q^a}{\pi v}\int_0^\infty
d\kappa \, \kappa J_0\left(\kappa\rho\right) \int_0^\infty\, d\omega
\frac{1}{k^2 \Delta\left(\omega,k\right)} \left[\cos \left \{
{\omega\left ( \frac{z}{v}
-t\right)}\right \}\,{\rm{Re}} \, \epsilon_L+
\sin \left \{\omega\left(\frac{z}{v}-t\right)\right \}\,{\rm{Im}}
\,\epsilon_L \right]
%\nonumber\\
%&=&\Phi^{a,R}\left(\rho,z,t\right)+\Phi^{a,I}
%\left(\rho,z,t\right)
\, . \label{potcyl}
\end{eqnarray}
where, $k=\sqrt{\kappa^2+{\omega^2}/{v^2}}$ and
$\Delta= \left({\rm {Re}}\, \epsilon_L\right)^2+
\left({\rm{Im}} \, \epsilon_L\right)^2$.

The potential in (\ref{pot1}) is also solved for two special cases, (a) {\it
 along the direction of motion of the parton $(\vec{r} \parallel \vec{v})$} and
(b) {\it  perpendicular to the direction of motion of the parton 
$(\vec{r} \perp \vec{v})$}. The potential for the parallel case is obtained as

\begin{equation}
\Phi^a_{\parallel}\left (\vec{r},\vec{v},t\right) = 
\frac{2Q^a}{\pi}\int_0^\infty dk \int_0^1\, dx \left[ 
\cos \left ( kx\left|\vec{\mathbf r}-\vec{\mathbf v}t\right|\right ) \,
\frac{ {\rm {Re}}\ \epsilon_L} {\Delta} +
\sin\left ( kx\left|\vec{\mathbf r}-\vec{\mathbf v}t\right|\right ) \,
\frac{ {\rm {Im}}\ \epsilon_L} {\Delta}\right] \ , \label{potpara}
\end{equation}

whereas that for the perpendicular case is

\begin{eqnarray}
\Phi^a_\bot \left(\vec{r},\vec{v},t\right) &=& \frac{Q^a}{2\pi^2}\int_0^\infty
dk\int_0^{2\pi}\,d\phi\int_{-1}^1 dx
\frac{e^{i k\left[r\cos{\phi}\sqrt{1-x^2}-vxt\right]}}{
\epsilon_L
\left(\omega=vkx,k\right)}\nonumber\\
&=& \frac{2Q^a}{\pi}\int_0^\infty dk\int_{0}^1 dx
\frac{J_0\left(kr\sqrt{1-x^2}\right)}{\Delta}\left[
\cos{(kvxt)} \, {\rm {Re}}\ \epsilon_L
- \sin{(kvxt)}\, {\rm {Im}}\ \epsilon_L \right]
\, , \label{potperp}
\end{eqnarray}
with $x=\cos \theta$. We point out here that some efforts were made
earlier in studying the wake potential~\cite{Chu89,Mustafa05} in a QGP.
However, those were incomplete, neglecting Mach cones and Cerenkov radiation 
in the wake of the plasma due to the motion of the charged particle.

\begin{figure}[!tbp]
%\begin{minipage}[h]{0.48\textwidth}
\begin{minipage}[t]{8cm}
\centering{\includegraphics[width=8cm,keepaspectratio]
{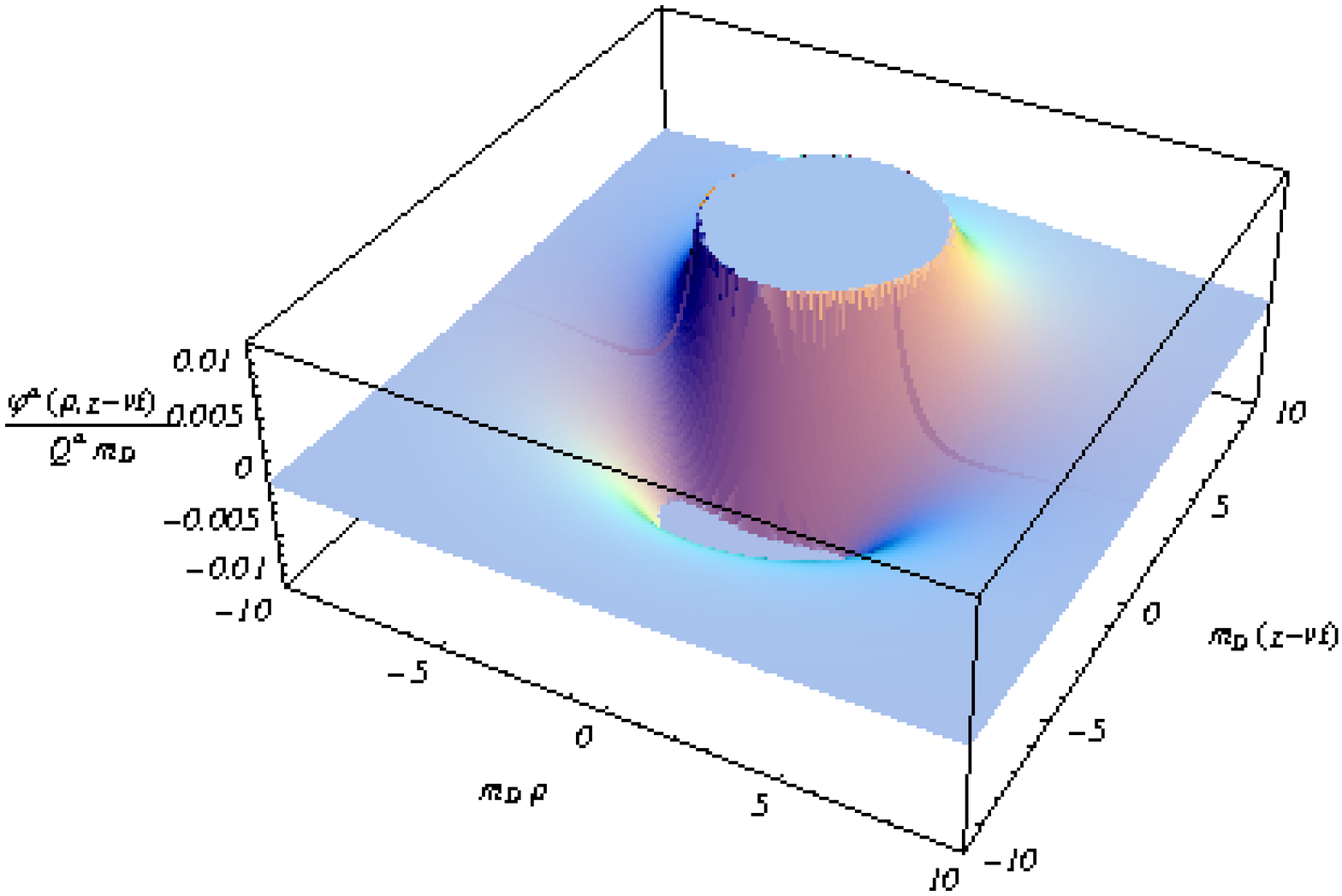}}
\end{minipage}
\hfill
\begin{minipage}[t]{8cm}
\centering{\includegraphics[width=8cm,keepaspectratio]
{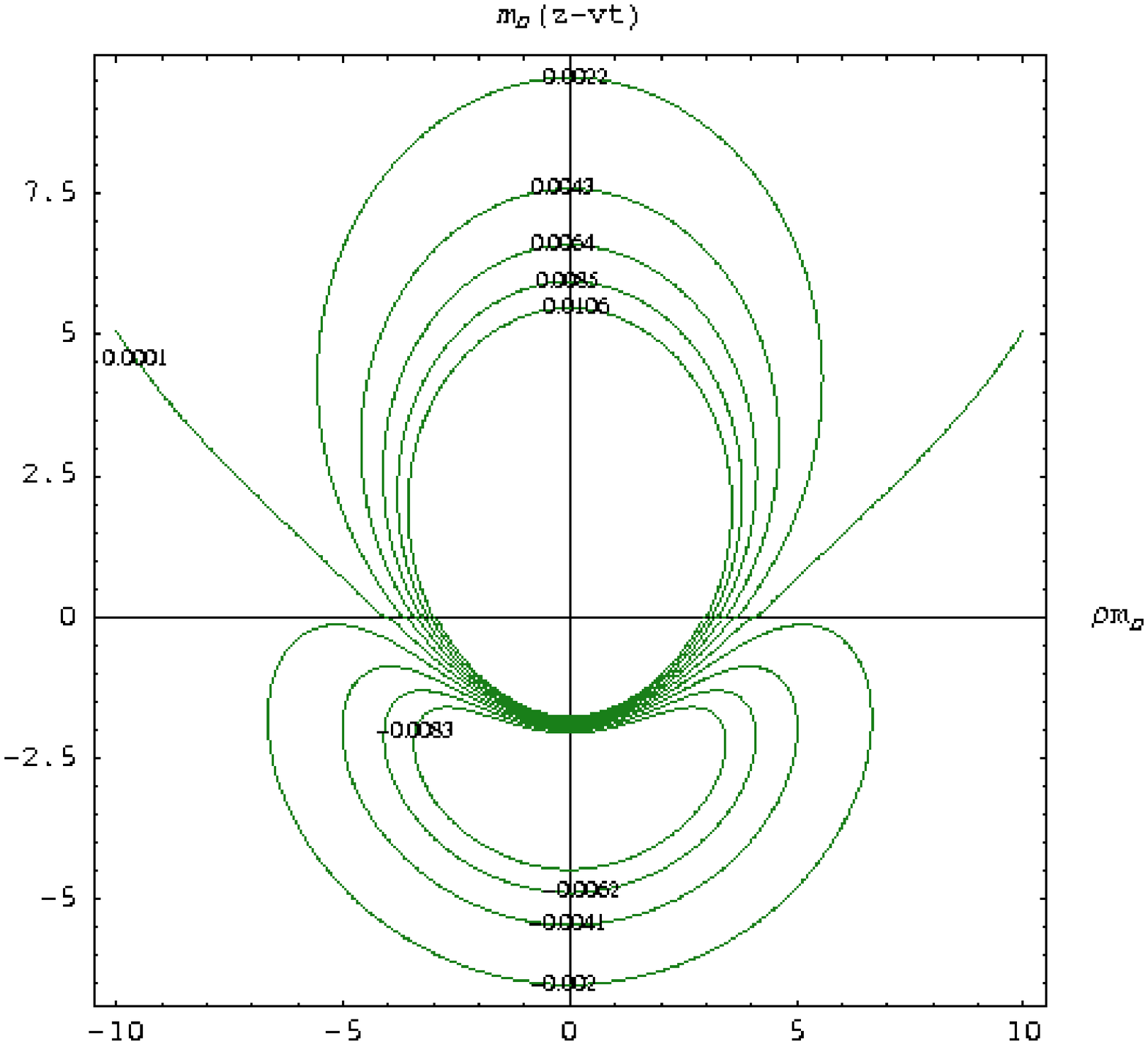}}
\end{minipage}
%\vspace*{-0.2in}
\caption{Left panel: Spatial distribution of the scaled wake potential 
with respect to $m_D$ from a jet 
with a fixed color charge $Q^a$ which is traveling with $v=0.55c$. 
Right panel: This plot shows the corresponding equipotential lines.}
\label{fig:pot_cp_v55}
\end{figure}

\begin{figure}[!tbp]
%\begin{minipage}[h]{0.48\textwidth}
\begin{minipage}[t]{8cm}
\includegraphics[width=8cm,keepaspectratio]
{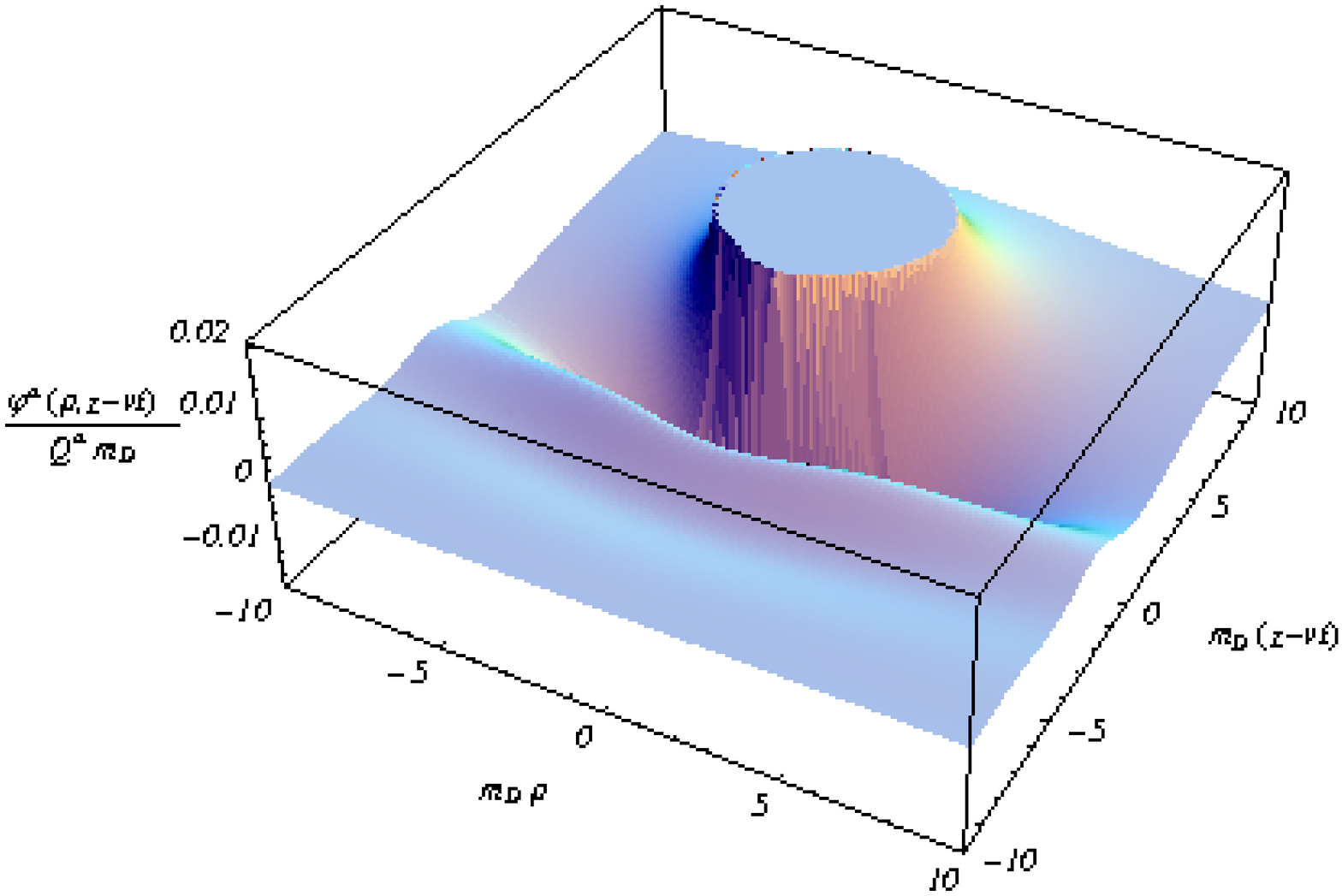}
\end{minipage}
\hfill
\begin{minipage}[t]{8cm}
\includegraphics[width=8cm,keepaspectratio]
{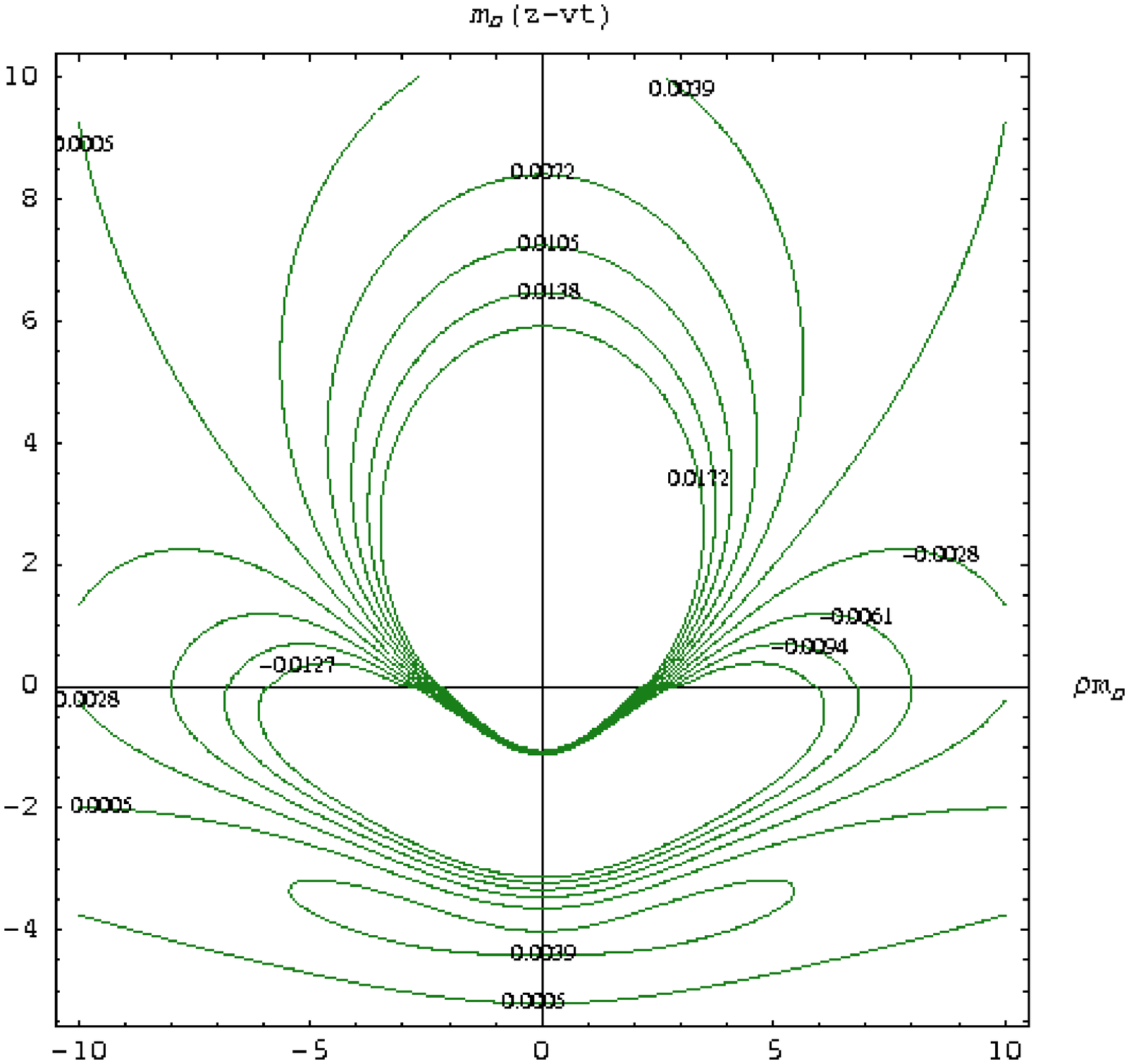}
\end{minipage}
%\vspace{0.8in}
\caption{Same as Fig.~\ref{fig:pot_cp_v55} but $v=0.99c$.} 
%Right panel: This plot shows the corresponding equipotential lines.}
\label{fig:pot_cp_v99}
\end{figure}

In Fig.~\ref{fig:pot_cp_v55} the wake potential is plotted for $v=0.55c$ in 
cylindrical coordinates which is scaled with $m_D$. (The wake potential 
is proportional to $m_D$.) 
The left panel shows a three dimensional plot
of the wake potential whereas the right panel displays the corresponding
equipotential lines.  It should be noted that the potential depends only 
on $v$ and not on $t$ as it should be for an isotropic and homogeneous plasma.
The spatial distribution of the potential shows the 
well known singular behavior at $r=0$, {\it i.e.} $z=0$ and $\rho=0$, which 
has been cut by hand through the scale restriction in the left panel.
It also exhibits a well defined negative minimum in the $(\rho - z)$ 
plane which has been formed due to the dynamical screening in the vicinity 
of the moving charged particle. This feature is also obvious
in the right panel where the equipotential lines are found to flip their sign
in the ($\rho - z$) plane corresponding to a moving induced charge cloud of 
opposite sign as shown in Fig.~\ref{fig:indchg_v55}, 
forming the wake around a moving particle.

When the velocity of the charged particle is higher than the average plasmon 
speed ($0.77c$), the induced charged cloud trailing the charged particle begins 
to oscillate which in turn indicates an oscillatory wake potential and has not
been found in earlier studies~\cite{Chu89,Mustafa05}.
In Fig.~\ref{fig:pot_cp_v99} we display the spatial distribution of the 
wake potential for  $v=0.99c$ which shows oscillations in the $(\rho - z$) 
plane. Such oscillations, as discussed in the preceding sub-sections, lead to 
an emission of Cerenkov like radiation and generate Mach shock waves.  
This is analogous to the wake phenomena in classical plasmas in the supersonic 
regime~\cite{Bashkirov04}.

\begin{figure}[!tbp]
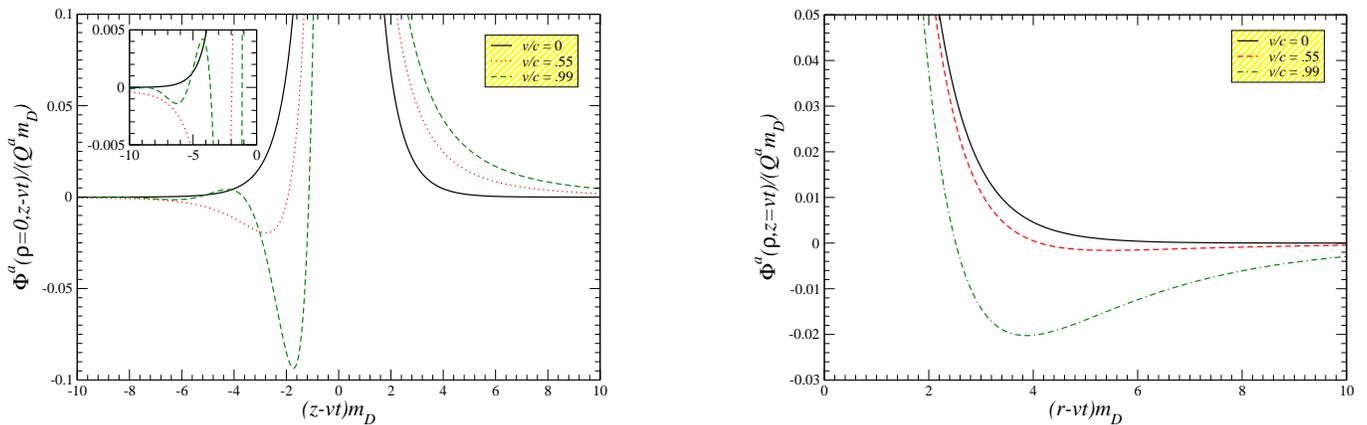

%\begin{minipage}[h]{0.4}
\vspace{0.5in}
\begin{minipage}[t]{8cm}
\centering{\includegraphics[width=8cm,keepaspectratio]
{paracomp_sc.eps}}
\end{minipage}
\hfill
\begin{minipage}[t]{8cm}
\centering{\includegraphics[width=8cm,keepaspectratio]
{perpcomp_sc.eps}}
\end{minipage}
\caption{Left panel: Scaled screening potential along the $z$-axis, {\it i.e.},
parallel to the direction of the moving color charge $Q^a$ for different velocities. 
Right panel: Same as left panel but along the $\rho$-axis, {\it i.e.},
perpendicular to the direction of the color charge.} 
\label{fig:para_v55}
\end{figure}

In Fig.~\ref{fig:para_v55} the wake potentials in the specific directions
corresponding to Figs.~\ref{fig:pot_cp_v55} and \ref{fig:pot_cp_v99}
are displayed. 
The left panel shows a wake potential along the $z$-axis, 
{\it i.e.}, parallel to the direction of the moving color charge. This could be directly
obtained from (\ref{potpara}) or by setting $\rho=0$ in (\ref{potcyl}).
As expected the singular behavior at $z=0$ is clearly reflected in this
plot. In the outward flow, {\it i.e.}, in the negative $z$-direction the
wake potential compared to the static one ({\it i.e.}, for $v=0$) falls 
off very fast, reverses its sign, exhibits a negative minimum and 
asymptotically approaches zero from below. For $v\leq 0.77c$ the potential 
in the outward flow is found to be of the Lennard-Jones type~\cite{Kittel} 
which 
has a short range repulsive part as well as a long range attractive part. 
With increase of $v$ the depth of the negative minimum increases and its 
position shifts towards the origin or to the particle.  When $0.77c < v <c$, 
the 
wake potential begins to oscillate. For $v=0.99c$ such oscillations in the 
wake potential along the direction of the motion are also clearly visible 
in the outward flow ($z<0$) from the inset of the left panel in 
Fig.~\ref{fig:para_v55}. 
On the other hand the potential for incoming flow\footnote{Our earlier 
study~\cite{Mustafa05} was restricted only to the outward flow and small 
particle velocities.}, {\it i.e.}, in the positive 
$z$-direction does not show any such structure and behaves more like a modified
Coulomb potential.  However, with the increase of $v$ it attains a Coulombic 
form indicating that the forward part of the screening cloud is not so strongly affected by the motion of the particle. 

Figs.~\ref{fig:pot_cp_v55}, \ref{fig:pot_cp_v99} and \ref{fig:para_v55} 
reveal that at finite $v$ the potential 
in the $z$-direction  becomes anisotropic and loses forward-backward symmetry 
with respect to the motion of the charged particle. The origin of this can be 
traced back to the second term in (\ref{potcyl}) and/or (\ref{potpara}) which 
is antisymmetric under inversion of $(z-vt)\rightarrow -(z-vt)$. This term
appears due to the antisymmetric nature of the imaginary part of 
$\epsilon_L(k,\omega)$ in (\ref{symdiel}). So, the effect of finite $v$ is 
not limited only to the deformation of the screening cloud, but, more 
importantly, it shifts the position of the test charge off the centre of the 
wake potential.
As a consequence, the test charge experiences a retarding force from the plasma.
This leads to a stopping power of the QGP against a moving test charge. The 
nature of the energy loss leading to such stopping power of the QGP has already
been discussed in the preceding subsec.~\ref{chg_ind}.  We also note that a 
minimum in the screening potential for a moving color 
charge was found in earlier works by Chu and Matsui~\cite{Chu89} and 
also by us~\cite{Mustafa05} in the direction of propagation. However, in 
Ref.~\cite{Chu89} Chu and Matsui did not report this minimum whereas our 
earlier study~\cite{Mustafa05} was incomplete concerning 
the forward-backward asymmetry of the wake structure and Cerenkov radiation. 

On the other hand, the right panel of Fig.~\ref{fig:para_v55} displays the
wake potential normal to the  direction of motion of the charged particle.
This could be obtained directly from (\ref{potperp}) or by setting $z=0$ 
in (\ref{potcyl}) which shows the usual singularity at $\rho=0$ and a 
symmetric behavior in $\rho$. It also falls off very fast compared to 
the static case and the nature of the potential is found to be 
of the Lennard-Jones type due to the deformed screening cloud forming the wake 
in the vicinity of the moving charged particle. With increase of $v$, the position
of the minimum shifts towards the origin and the depth of it increases. Such kind 
of potential in the normal direction  implies that a part of 
the screening cloud forming the wake due to the motion of the charged particle 
along the $z$-direction is also moving away in the perpendicular direction, 
resulting in a transverse flow in the medium. This behavior was not observed
in earlier studies~\cite{Chu89,Mustafa05}.

The negative minimum in the wake potential indicates
an induced space charge density of opposite sign. Thus, a particle moving
relative to a particle in the induced space charge density would constitute
a dipole oriented along the direction of motion. In the next sub-section,
we calculate the potential due to such dipole interaction in the QGP.

\subsection{Dipole Potential in the QGP}
\label{dipole_pot}

We consider two color charges $Q^a$ and $Q^b$ separated by a distance $r$.
The change in free energy of the system to bring the two widely separated 
color charges together~\cite{Lebellac,Blaizot,Zee} is
\begin{eqnarray}
\Delta {\mathcal F} &=& \frac{1}{2} \int d^3 {\vec {\mathbf x}} \, 
J^\mu_{\rm{ext}}\left( x \right)A_\mu\left( x\right )
= \frac{1}{2} \int \frac{d^3 k}{\left(2\pi\right)^3} \int_{-\infty}^\infty
\frac{d\omega}{\left(2\pi\right)} \int_{-\infty}^\infty
\frac{d\omega^\prime}{\left(2\pi\right)}
e^{it\left(\omega+\omega^\prime\right)}J^\mu_{\rm{ext}}\left(\omega,-
\vec{\mathbf k}\right) A_\mu \left(\omega^\prime,\vec{\mathbf k}\right)
\, , \label{freeenergy1}
\end{eqnarray}
where $ J^\mu_{\rm{ext}}\left( x \right) = \left(\rho_{\rm{ext}},
\vec{\mathbf J}_{\rm{ext}} \right )$ is the sum of the two external currents, 
 \( J^\mu_{\rm{ext}}\left( x \right) 
= J^\mu_1\left( x \right)+J^\mu_2\left( x \right)\), and 
$A_\mu\left({\vec {\mathbf x}}\right)$ is the
associated gauge field. The entropy generation is neglected in (\ref{freeenergy1}). 
%The factor $1/2$ is to correct the double counting.

We assume that the average value of $A_\mu$ vanishes in 
equilibrium \(\left\langle A_\mu \right\rangle_{\mbox{eq}}=0\). 
The induced expectation value of the vector potential, $A_\mu$
follows from linear response theory~\cite{Weldon82} as
\begin{equation}
\label{expea}
\langle A_\mu\rangle_\beta=4\pi\Delta_{\mu\nu}
\left(\Omega,{\vec {\mathbf k}}\right)
\, J^{\nu}_{\rm{ext}}(K)\ \ , 
\end{equation}
where, $\Omega=u.K, \ k=\sqrt{\Omega^2-K^2}$ are the Lorentz invariant energy
and 3-momentum, respectively. $\Delta_{\mu\nu}$ is the propagator of
the gauge boson exchanged between the two currents which is given in 
(\ref{fullprop}). Combining (\ref{expea}) with (\ref{freeenergy1}) we can 
write 
\begin{eqnarray}
\Delta {\mathcal F} &=& 
 2\pi \int \frac{d^3 k}{\left(2\pi\right)^3} \int_{-\infty}^\infty
\frac{d\omega}{\left(2\pi\right)} \int_{-\infty}^\infty
\frac{d\omega^\prime}{\left(2\pi\right)}
e^{it\left(\omega+\omega^\prime\right)}
J^\mu_{\rm{ext}}\left(\omega,-
\vec{\mathbf k}\right) \Delta_{\mu\nu}(\omega^\prime,{\vec{\mathbf k}})
J^\nu_{\rm{ext}}(\omega^\prime,{\vec{\mathbf k}}) 
\, . \label{freeenergy2}
\end{eqnarray}
One can obtain the dipole interaction in (\ref{freeenergy2}) by
separating~\cite{Weldon82} the current\footnote{Henceforth, 
we drop the suffix {\bf `ext'} from the external current for
convenience.}, in general, into a charge density that 
moves with the fluid velocity $u^\mu$ and a spacelike current flow
either longitudinal or transverse to ${\tilde K}^\mu$ as defined in 
(\ref{orthotv}):
\begin{equation}
J^\mu(K)=u^\mu u^\alpha J_\alpha(K)+J^\mu_{L}(K)+J^\mu_T(K) \ \ ,
\label{gencur}
\end{equation} 
where
\begin{eqnarray}
J^\mu_L (K)&=& \frac{\tilde{K}^\mu \omega}{k^2} u^\alpha J_\alpha(K) \, ,
\nonumber \\
J^\mu_T(K)&=&\left [\tilde{\eta}^\mu_\lambda + \frac{\tilde{K}^\mu 
\tilde{K}_\lambda}{k^2}\right ] J^\lambda(K) \, . \label{decomgc}
\end{eqnarray}
Using (\ref{fullprop}) and (\ref{gencur})
the dipole interaction can be obtained as
\begin{eqnarray}
J^\mu\left(\omega,-
\vec{\mathbf k}\right) \Delta_{\mu\nu}(\omega^\prime,{\vec{\mathbf k}})
J^\nu (\omega^\prime,{\vec{\mathbf k}})&=& 
%\left[u\cdot J(\omega,-\vec{\mathbf k}) \right]
%\left[u\cdot J(\omega^\prime,\vec{\mathbf k}) \right]+
-\frac{k^2+\omega\omega^\prime}{\omega\omega^\prime}
 \frac{
\vec{\mathbf J}_L(\omega,-\vec{\mathbf k}) {\mathbf \cdot}
\vec{\mathbf J}_L(\omega,\vec{\mathbf k})} 
{{\omega^\prime}^2-k^2-\Pi_L(\omega^\prime, k)}
 - \frac{\vec{\mathbf J}_T(\omega,-\vec{\mathbf k}) {\mathbf \cdot}
\vec{\mathbf J}_T(\omega,\vec{\mathbf k})} 
{{\omega^\prime}^2-k^2-\Pi_T(\omega^\prime,k)} \nonumber \\
&=&
\frac{
\vec{\mathbf J}(\omega,-\vec{\mathbf k}) {\mathbf \cdot}
\vec{\mathbf J}(\omega,\vec{\mathbf k})+
\frac{\omega\omega^\prime}{k^2}
{\rho(\omega,-\vec{\mathbf k})\rho(\omega,\vec{\mathbf k})} } 
{{\omega^\prime}^2-k^2-\Pi_T(\omega^\prime, k)} \, \,
\nonumber \\
&& - \frac{k^2+\omega\omega^\prime}{k^2}
\frac{\rho(\omega,-\vec{\mathbf k})\rho(\omega,\vec{\mathbf k})} 
{{\omega^\prime}^2-k^2-\Pi_L(\omega^\prime, k)} \, \, , 
\label{dipint}
\end{eqnarray}
where $u.J=\rho$. Though the gluon propagator in (\ref{fullprop}) is
gauge dependent, the dipole interaction becomes gauge independent.

We begin with the known example of screening of two static charges $Q^a$ and
$Q^b$ at  position $\vec{\mathbf x}_1$ and $\vec{\mathbf x}_2$, 
respectively. The corresponding dipole current is given by
\begin{equation}
\label{statini}
J^\mu\left(x\right) =J^\mu_1\left(x\right)+J^\mu_2\left(x\right) =
\delta^\mu_0\left[ Q^a\delta \left(\vec{\mathbf x}-\vec{\mathbf x}_1\right)
+Q^b\delta \left(\vec{\mathbf x}-
\vec{\mathbf x}_2\right)\right] \stackrel{\mbox{FT}}{=} 
2\pi \delta_0^\mu \delta
\left(\omega\right)\left[
Q^ae^{-i\vec{\mathbf k}\cdot \vec{\mathbf x}_1}+Q^be^{-i\vec{\mathbf k}\cdot
\vec{\mathbf x}_2}\right]\,.
\end{equation}
Combining (\ref{statini}), (\ref{dipint}) and (\ref{freeenergy2}), and also
dropping the self coupling terms we get the well known Yukawa potential as
\begin{equation}
\Delta{\mathcal{F}}^{ab}(r) = \frac{Q^aQ^b}{2\pi^2}\int d^3k\,
\frac{ \cos \left ( {\vec{\mathbf k}} \cdot {\vec{\mathbf r}}\right )}
{\kappa^2+k^2_z+m_D^2} =
Q^aQ^b\,\frac{e^{-m_D r}}{r}\, , \label{dipoyuka}
\end{equation}
where, $\vec{\mathbf r} ={\vec{\mathbf x_1}}
-{\vec{\mathbf x_2}}$ and ${\vec{\mathbf r}} = \left(\rho,0,z\right)$.
It is also noteworthy to mention that in Ref.~\cite{Gale87} a static 
screening potential was also obtained in a QGP, where a polarization tensor
beyond the high temperature limit was used. However, this approach in 
Ref.~\cite{Gale87} has its limitation because a gauge dependent and 
incomplete approximation for the polarization tensor was used.
                                                                                
Now for two comoving charges $Q^a$ and $Q^b$ the dipole current 
can be written as
%follows 
%from (\ref{initchgj})
\begin{eqnarray}
J_\mu\left(t,{\vec{\mathbf x}}\right) = \left(1,{\vec{\mathbf v}}\right)
\left[Q^a \delta\left({\vec{\mathbf x}}-{\vec{\mathbf x}_1}-
{\vec{\mathbf v}}t
\right)+Q^b \delta\left({\vec{\mathbf x}}-{\vec{\mathbf x}_2}-
{\vec{\mathbf  v}}t\right)\right]
\stackrel{\mbox{FT}}{=} 2\pi\left(1,{\vec{\mathbf v}}\right) 
\delta\left(\omega
-{\vec{\mathbf k}}\cdot{\vec{\mathbf v}}\right)
\left[Q^a e^{-i{\vec{\mathbf k}}\cdot{\vec{\mathbf x}_1}}
+Q^b e^{-i{\vec{\mathbf k}}\cdot{\vec{\mathbf x}_2}}\right]\, ,
\label{dipoinit}
\end{eqnarray}
%Substituting (\ref{dipoinit}) in (\ref{freeenergy1}) using (\ref{fullprop}) 
and we obtain the dipole potential as 
\begin{eqnarray}
\Delta{\mathcal {F}}^{ab}(r;\rho,z) &=& \frac{Q^aQ^b}{2\pi^2}\int d^3k \,
\cos\left( {{\vec{\mathbf k}} \cdot{\vec{\mathbf r}}} \right )
\left[\frac{v^2-\frac{\omega^2}{k^2}}
{K^2-\Pi_T\left(\omega,k\right)}-
\frac{1-\frac{\omega^2}{k^2}}
{K^2 -\Pi_L\left(\omega,k\right)}\right]_{\omega={\vec{\mathbf k}}
\cdot {\vec{\mathbf v}}}\,,\nonumber \\
&=& \frac{2Q^aQ^b}{\pi}
\int_0^\infty d\kappa\, \kappa J_0\left(\kappa\rho\right)
\int_0^\infty dk_z\,
\cos\left({k_z z}\right) \,\, \left\{{\rm{Re}} \, 
\left[\frac{v^2-\frac{\omega}{k^2}}
{K^2-\Pi_T\left(\omega,k\right)}-
\frac{1-\frac{\omega^2}{k^2}}
{K^2-\Pi_L\left(\omega,k\right)}\right]_{\omega={\vec{\mathbf k}}
\cdot {\vec{\mathbf v}}}
\right\}.\label{eqn7}
\end{eqnarray}

The following features are transparent in (\ref{eqn7}). The first term 
corresponds to the transverse (magnetic) interaction whereas the second term 
is due to the longitudinal (electric) interaction\footnote{In our earlier 
study~\cite{Mustafa05} the two body potential was obtained by averaging the
single body potential and found to depend only on the longitudinal (electric)
interaction.}. Like the single body potential in 
(\ref{potcyl}) the two body potential is also symmetric in $\rho$.  Moreover, 
in (\ref{eqn7}) both imaginary parts corresponding to the longitudinal and
transverse response functions drop out because they are odd in $k_z$, 
which make both terms symmetric under the inversion of $z\ \rightarrow \ -z$,
thus the total potential. 
This is, however, unlike the single body potential in (\ref{potcyl}) where 
the presence of an imaginary part of the longitudinal response 
function was found to be responsible for the asymmetric behavior. 
In addition the two body potential should be symmetric in $z$ as the
two color charges are moving opposite to each other which is obvious from
the dipole interaction in (\ref{dipint}). Recalling Ampere's law this would
also amount to an  attraction between color charges traveling opposite to 
each other along the $z$ axis. Both the electric and magnetic interactions play
a crucial role, as we will see below.

\begin{figure}[!tbp]
\vspace{.5cm}
%\begin{minipage}[h]{0.48\textwidth}
\begin{minipage}[t]{8cm}
\includegraphics[width=8cm,keepaspectratio]
{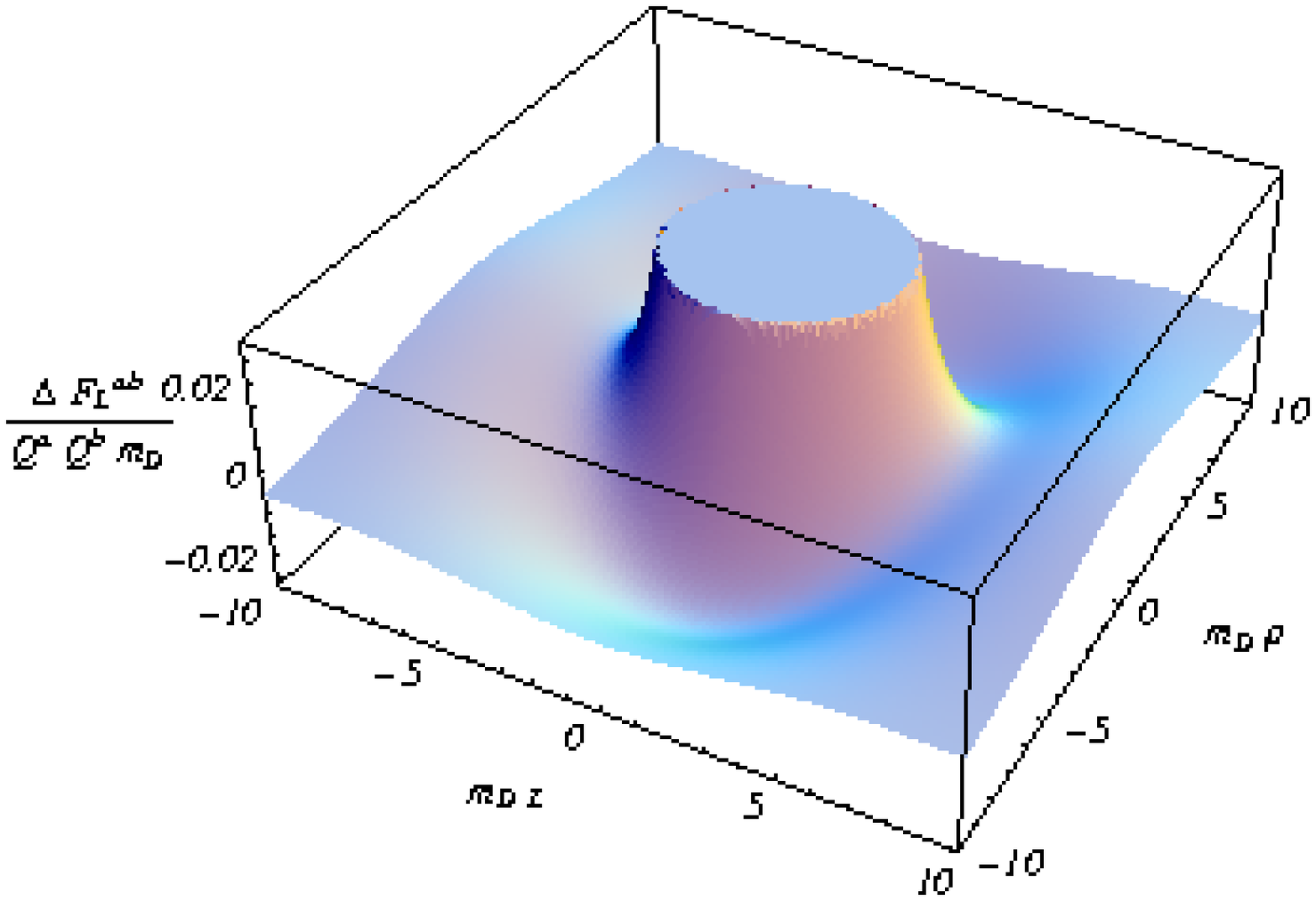}
\end{minipage}
\hfill
\begin{minipage}[t]{8cm}
\includegraphics[width=8cm,keepaspectratio]
{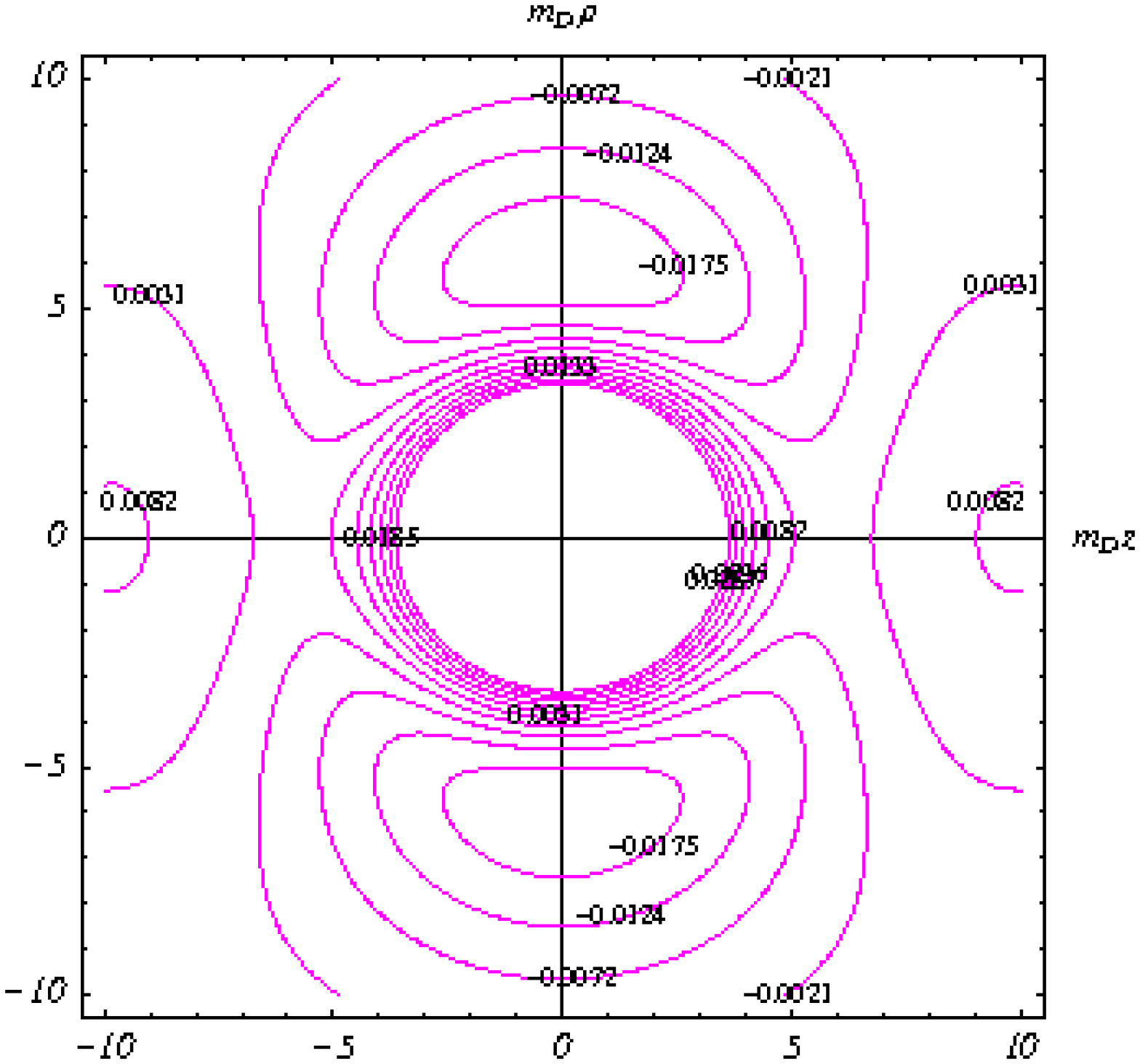}
\end{minipage}
%\vspace{0.8in}
\caption{Left panel: Spatial distribution of the longitudinal part of the 
scaled dipole potential with respect to $m_D$ in which the two color charges 
are traveling antiparallel to each other with $v=0.55c$. 
Right panel: This plot shows the corresponding equipotential lines.}
\label{fig:long_v55}
\end{figure}

\begin{figure}[!tbp]
\vspace{.5cm}
%\begin{minipage}[h]{0.48\textwidth}
\begin{minipage}[t]{8cm}
\includegraphics[width=8cm,keepaspectratio]
{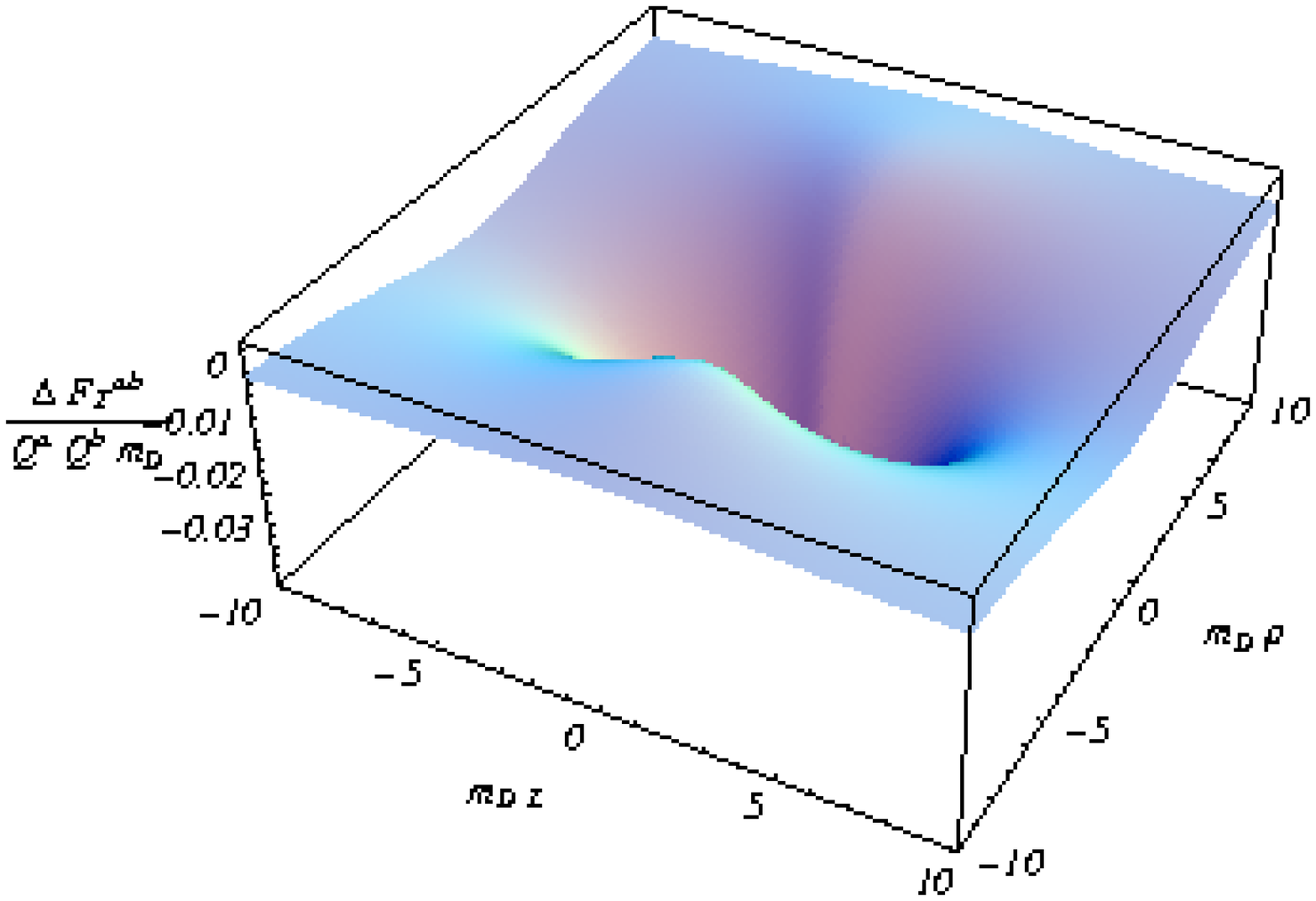}
\end{minipage}
\hfill
\begin{minipage}[t]{8cm}
\includegraphics[width=8cm,keepaspectratio]
{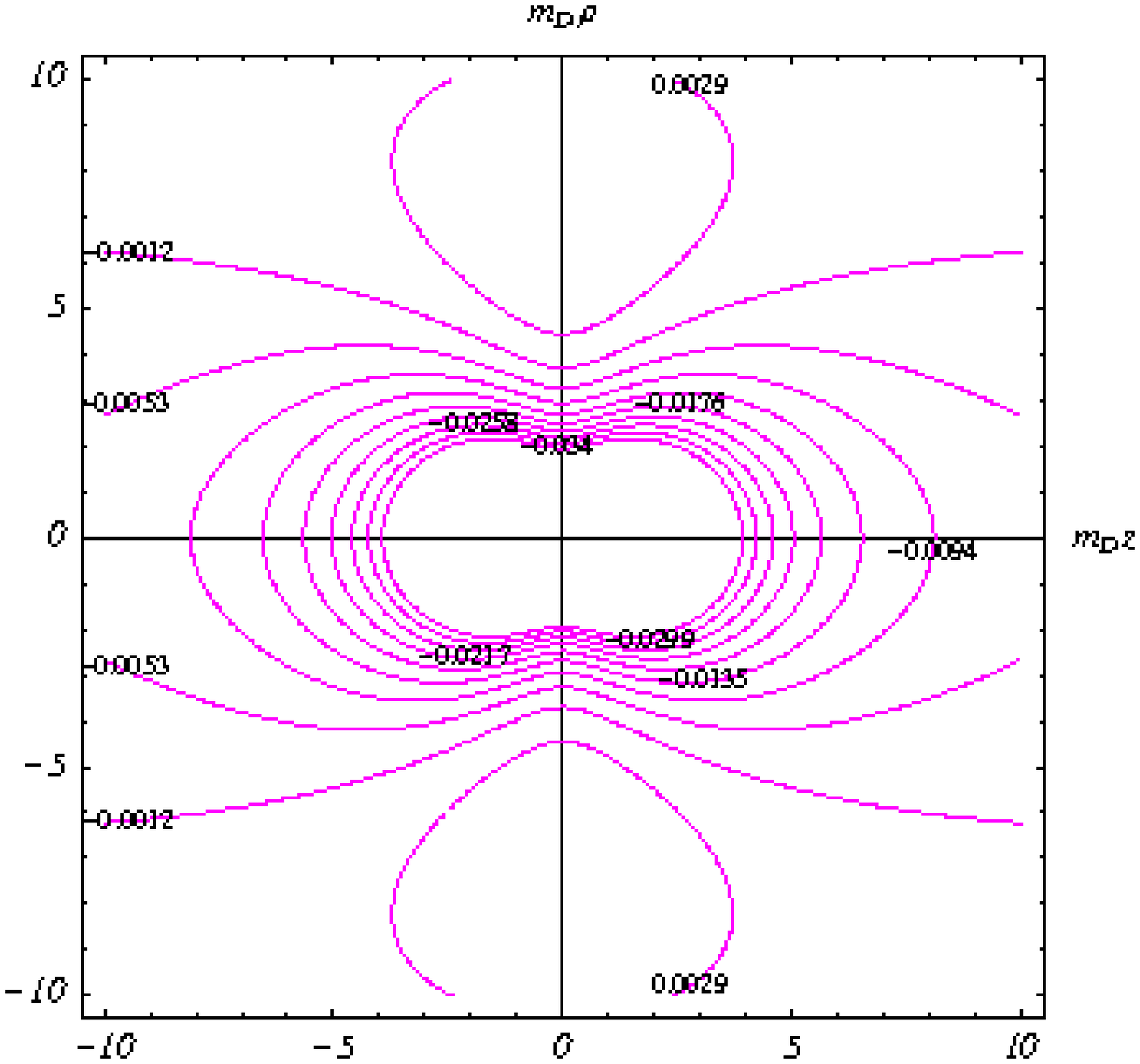}
\end{minipage}
%\vspace{0.8in}
\caption{Same as Fig.~\ref{fig:long_v55} but showing the transverse part.}
%Left panel: Spatial distribution of the scaled dipole potential  
%with respect to $m_D$ in which two color charges are traveling antiparallel 
%to each other with $v=0.55c$. 
%Right panel: This plot shows the corresponding equipotential lines.}
\label{fig:trans_v55}
\end{figure}

\begin{figure}[!tbp]
\vspace{.6cm}
%\begin{minipage}[h]{0.48\textwidth}
\begin{minipage}[t]{8cm}
\includegraphics[width=8cm,keepaspectratio]
{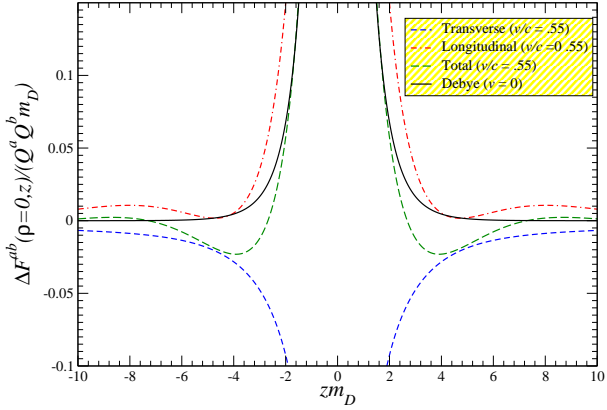}
\end{minipage}
\hfill
\begin{minipage}[t]{8cm}
\includegraphics[width=8cm,keepaspectratio]
{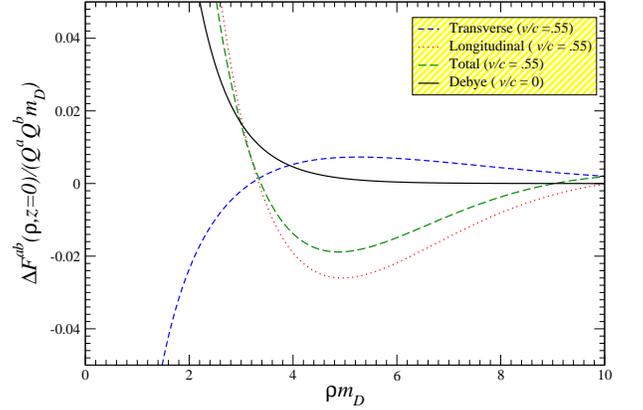}
\end{minipage}
%\vspace{0.8in}
\caption{Left panel: Scaled two body potential along the direction of motion 
when two color charges are traveling antiparallel to each other with
$v=0.55c$.  
Right panel: Same as left panel but normal to the direction of motion.}
\label{fig:specific_v55}
\end{figure}

In Fig.~\ref{fig:long_v55} the spatial distribution of the 
scaled~\footnote{The potential is scaled with $m_D$ as well as
with the interaction strength, $Q^aQ^b$. So, the details of the potential 
will depend on the temperature, the strong coupling constant, and the sign 
of the interaction strength.} 
longitudinal 
part of the two body potential in cylindrical coordinate is displayed for 
$v=0.55c$. The left panel shows a three dimensional plot of it whereas the 
right panel displays the corresponding equipotential lines. 
The longitudinal part has a singularity at $r=0$, {\it i.e.}, $z=0$ and 
$\rho=0$, clearly reflected in Fig.~\ref{fig:long_v55}.
As discussed above the potential plotted in both panels is found to be 
completely symmetric in the ($\rho -z$) plane along with a variable minimum.
Such a minimum in the $(\rho -z$) plane is due to the fact that the electric
interaction contributes differently in $z$ as well as in $\rho$ direction.
The potential has a positive minimum in the longitudinal direction, {\it i.e.},
along the $z$ 
axis and then tends to zero for large $|z|$. This can also be seen from 
the left panel of Fig.~\ref{fig:specific_v55}, which can be obtained from
(\ref{eqn7}) by setting $\rho=0$. On the other hand, in the transverse 
direction, {\it i.e.}, along the $\rho$ direction, the electric contribution
has a  well defined negative minimum, it also  attains a positive maximum at
large $\rho$ and slowly approaches zero asymptotically. This feature
can clearly be seen from  the right panel of Fig.~\ref{fig:specific_v55},
which is obtained by solving (\ref{eqn7}) with $z=0$. 

The contribution of the scaled transverse part is shown in 
Fig.~\ref{fig:trans_v55}
which is also found to be symmetrical in the $(\rho -z)$ plane along with a well 
defined negative minimum. However, both panels indicate that the magnetic 
interaction contributes differently in $z$ as well in $\rho$ directions, 
which can also be seen from Fig.~\ref{fig:specific_v55}.
In the limit $v\rightarrow 0$, the transverse part vanishes. However, for any 
nonzero value of $v$ the transverse part begins with a finite negative value
for both cases. As $|z|$ increases, its magnitude gradually decreases and 
approaches zero for large $|z|$ (left panel of Fig.~\ref{fig:specific_v55}) 
whereas in the transverse direction its magnitude decreases faster,  becomes 
positive at certain values of $\rho$, and then slowly approaches zero for 
large values of $\rho$ (right panel of Fig.~\ref{fig:specific_v55}).

\begin{figure}[!tbp]
\vspace{.5cm}
%\begin{minipage}[h]{0.48\textwidth}
\begin{minipage}[t]{8cm}
\includegraphics[width=8cm,keepaspectratio]
{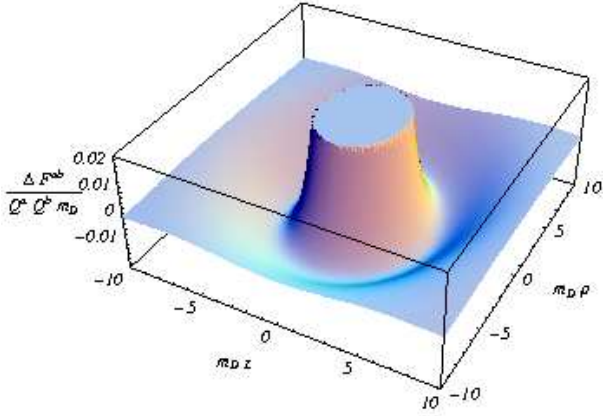}
\end{minipage}
\hfill
\begin{minipage}[t]{8cm}
\includegraphics[width=8cm,keepaspectratio]
{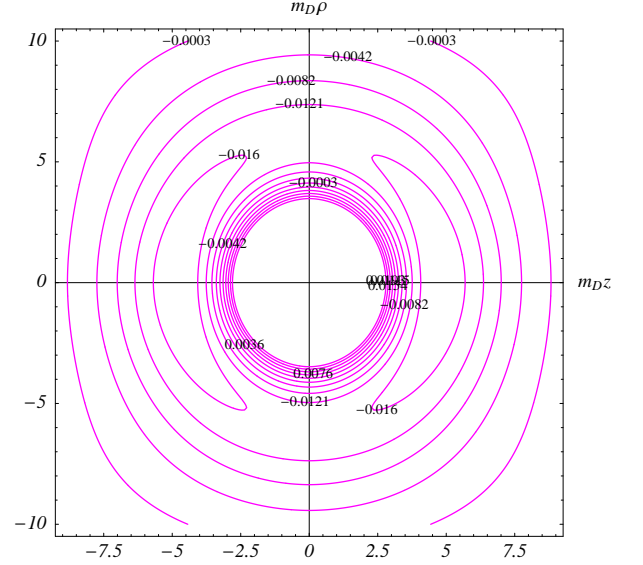}
\end{minipage}
%\vspace{0.8in}
\caption{Total contribution for $v=0.55c$}  
\label{fig:total_v55}
\end{figure}

\begin{figure}[!tbp]
\vspace{.5cm}
%\begin{minipage}[h]{0.48\textwidth}
\begin{minipage}[t]{8cm}
\includegraphics[width=8cm,keepaspectratio]
{2b_v55_long.eps}
\end{minipage}
\hfill
\begin{minipage}[t]{8cm}
\includegraphics[width=8cm,keepaspectratio]
{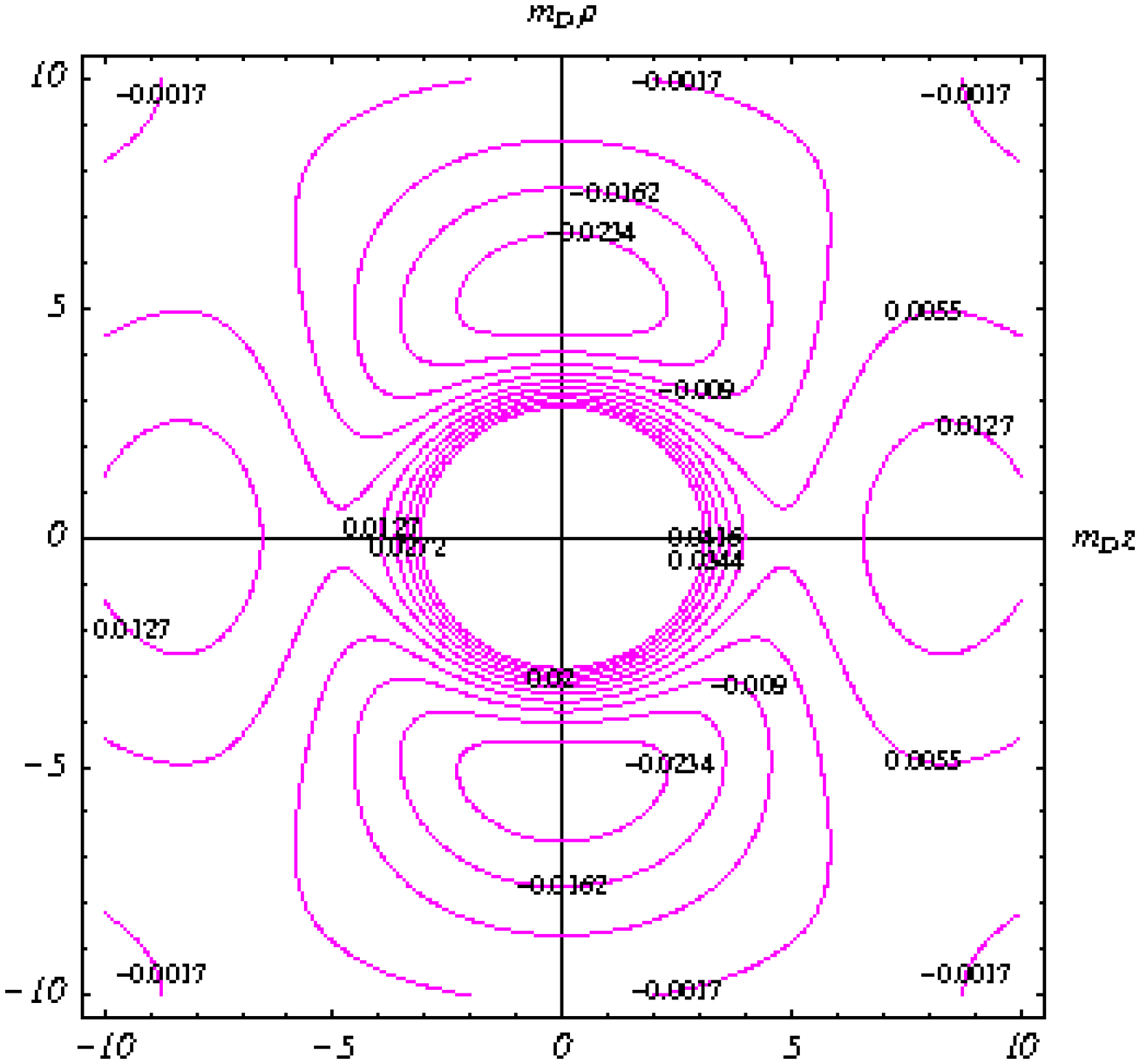}
\end{minipage}
%\vspace{0.8in}
\caption{Left panel: Spatial distribution of the longitudinal part of the 
scaled dipole potential with respect to $m_D$ in which two color charges 
are traveling antiparallel to each other with $v=0.99c$. 
Right panel: This plot shows the corresponding equipotential lines.}
\label{fig:long_v99}
\end{figure}

\begin{figure}[!tbp]
\vspace{.6cm}
%\begin{minipage}[h]{0.48\textwidth}
\begin{minipage}[t]{8cm}
\includegraphics[width=8cm,keepaspectratio]
{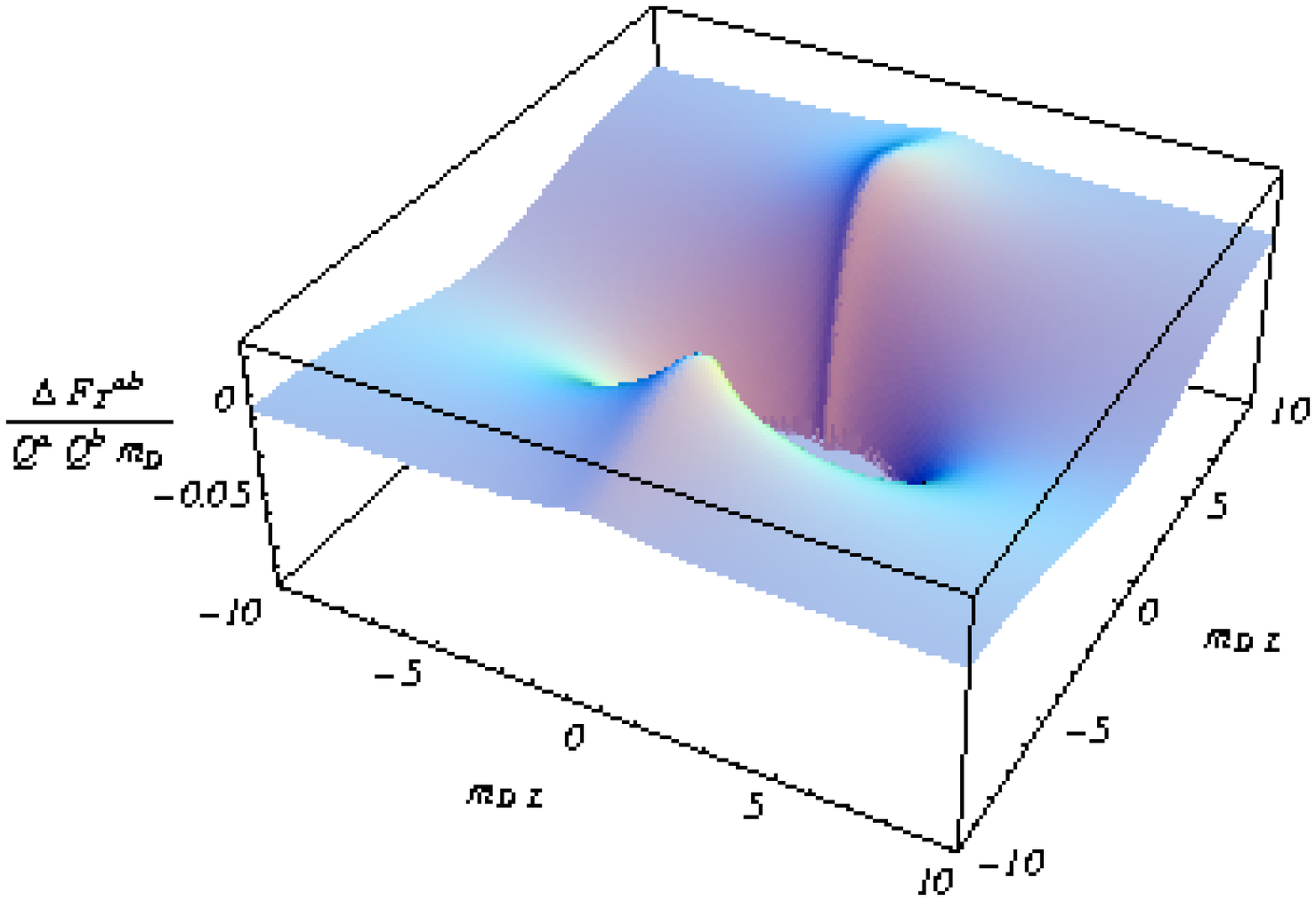}
\end{minipage}
\hfill
\begin{minipage}[t]{8cm}
\includegraphics[width=8cm,keepaspectratio]
{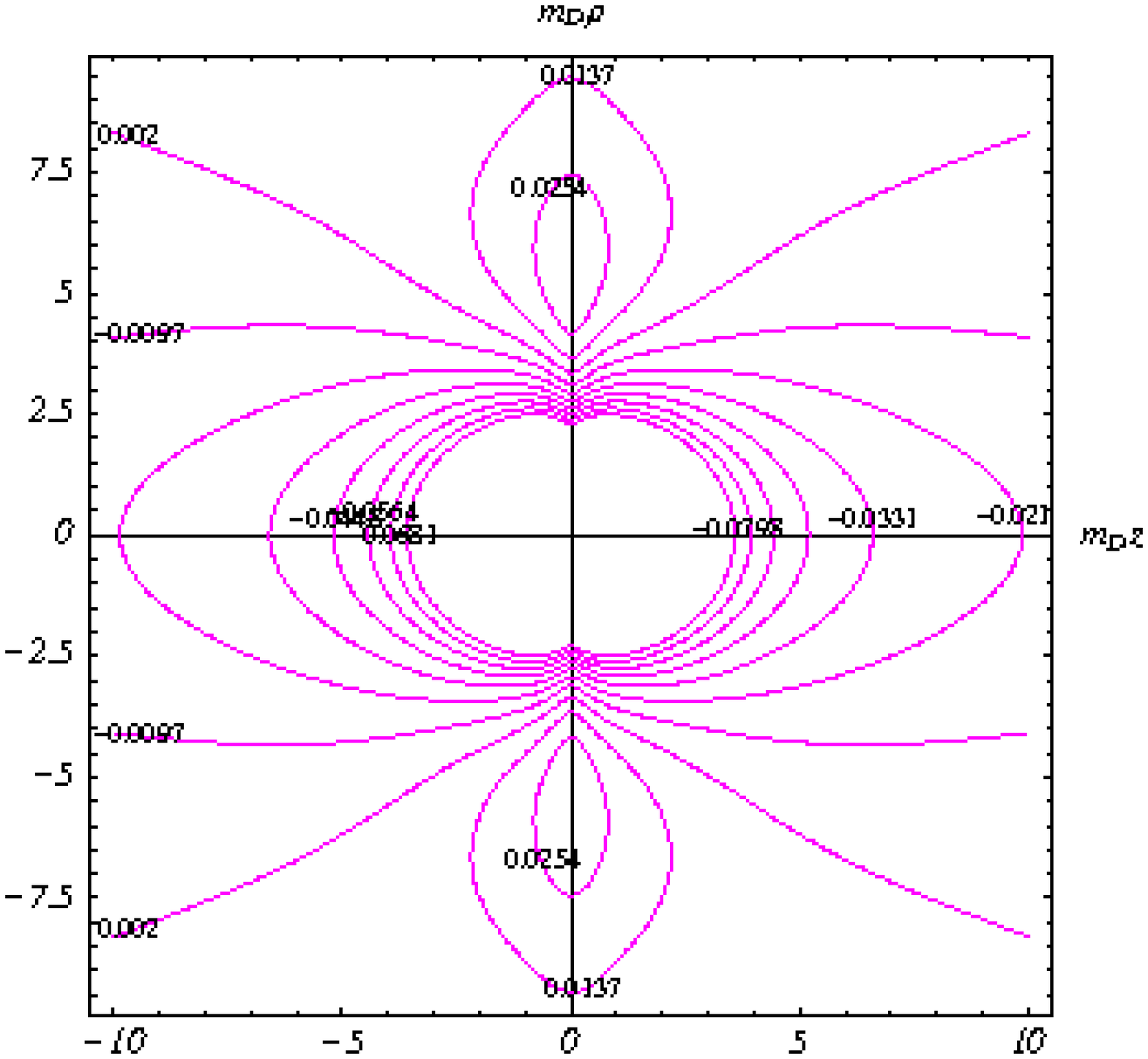}
\end{minipage}
%\vspace{0.8in}
\caption{Same as Fig.~\ref{fig:long_v99} but showing the transverse part.}
\label{fig:trans_v99}
\end{figure}
The spatial distribution of the total potential for $v=0.55c$ is displayed in 
Fig.~\ref{fig:total_v55}, which appears due to the compensating effects 
between the electric and the magnetic interactions as discussed above. The 
resulting potential shows the usual singularity of the screening potential at 
$r=0$ ($z=0$ and $\rho =0$) and a completely symmetric behavior along with a 
pronounced negative minimum in the $(\rho -z)$ plane, which are clearly 
reflected in both panels of Fig.~\ref{fig:total_v55}. However, the detailed 
features in the specific direction can also be seen from both panels of 
Fig.~\ref{fig:specific_v55}. The potential along the $|z|$-direction as 
shown in the left panel falls off like that of a static one, flips its sign, 
exhibits a negative minimum, and then oscillates around zero. As can be seen
from the left panel, the competing effects between electric and magnetic
contributions with opposite sign result in an oscillatory potential along
the $z$ direction. 
On the other hand, the potential in the transverse direction (right panel) 
falls off slowly compared to the static one, flips its sign, and 
exhibits a negative minimum at some values of $\rho$,  again changes its sign
to attain small positive maximum and then tends to zero at large $\rho$.
It can be seen that the nature of the potential is completely dictated by 
the electric interaction rather than by the magnetic one. In both directions there are
oscillations at large distances but the form of the potential 
mostly resembles the Lennard-Jones type with a pronounced repulsive 
as well attractive part.

\begin{figure}[!tbp]
\vspace{.5cm}
%\begin{minipage}[h]{0.48\textwidth}
\begin{minipage}[t]{8cm}
\includegraphics[width=8cm,keepaspectratio]
{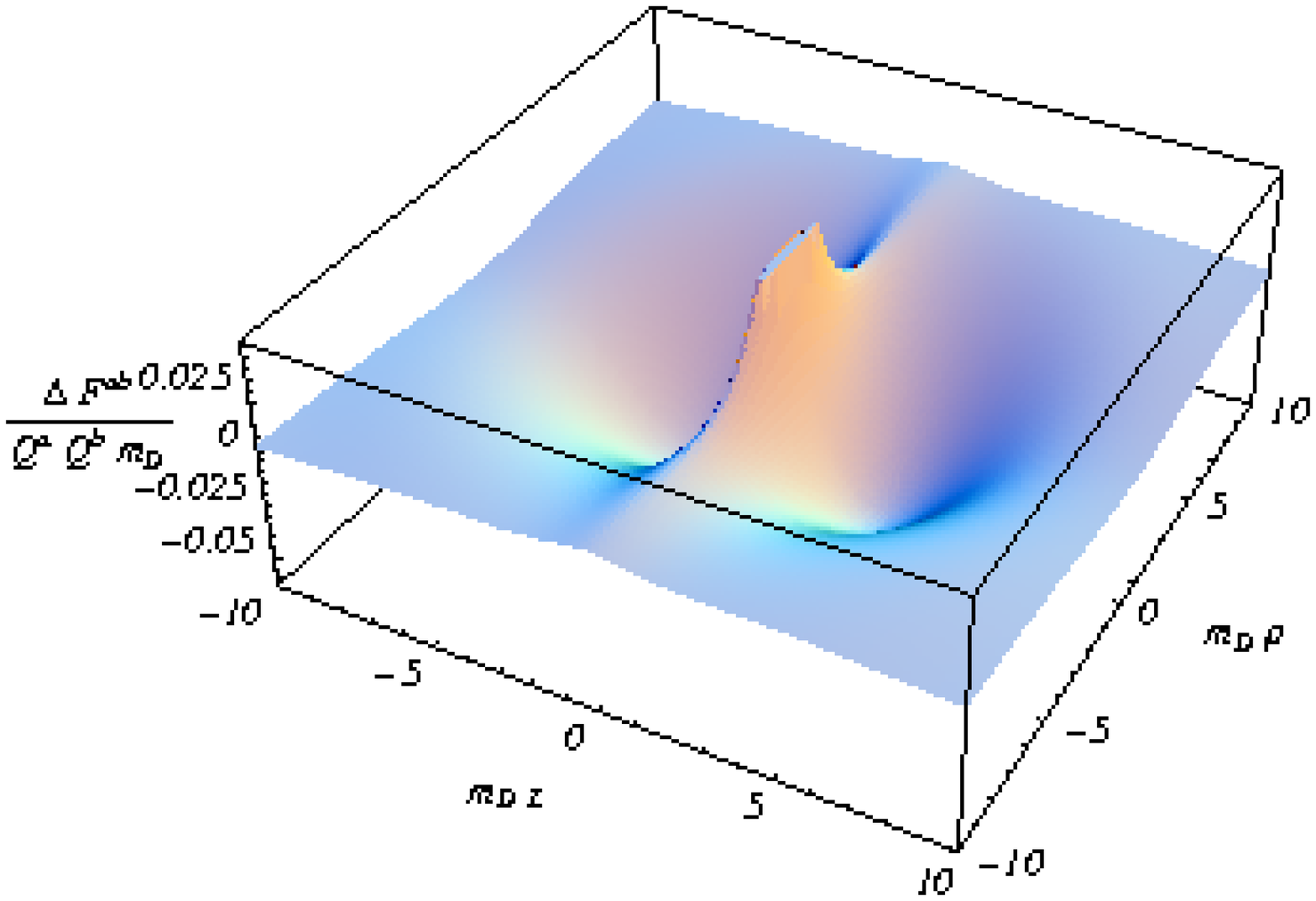}
\end{minipage}
\hfill
\begin{minipage}[t]{8cm}
\includegraphics[width=8cm,keepaspectratio]
{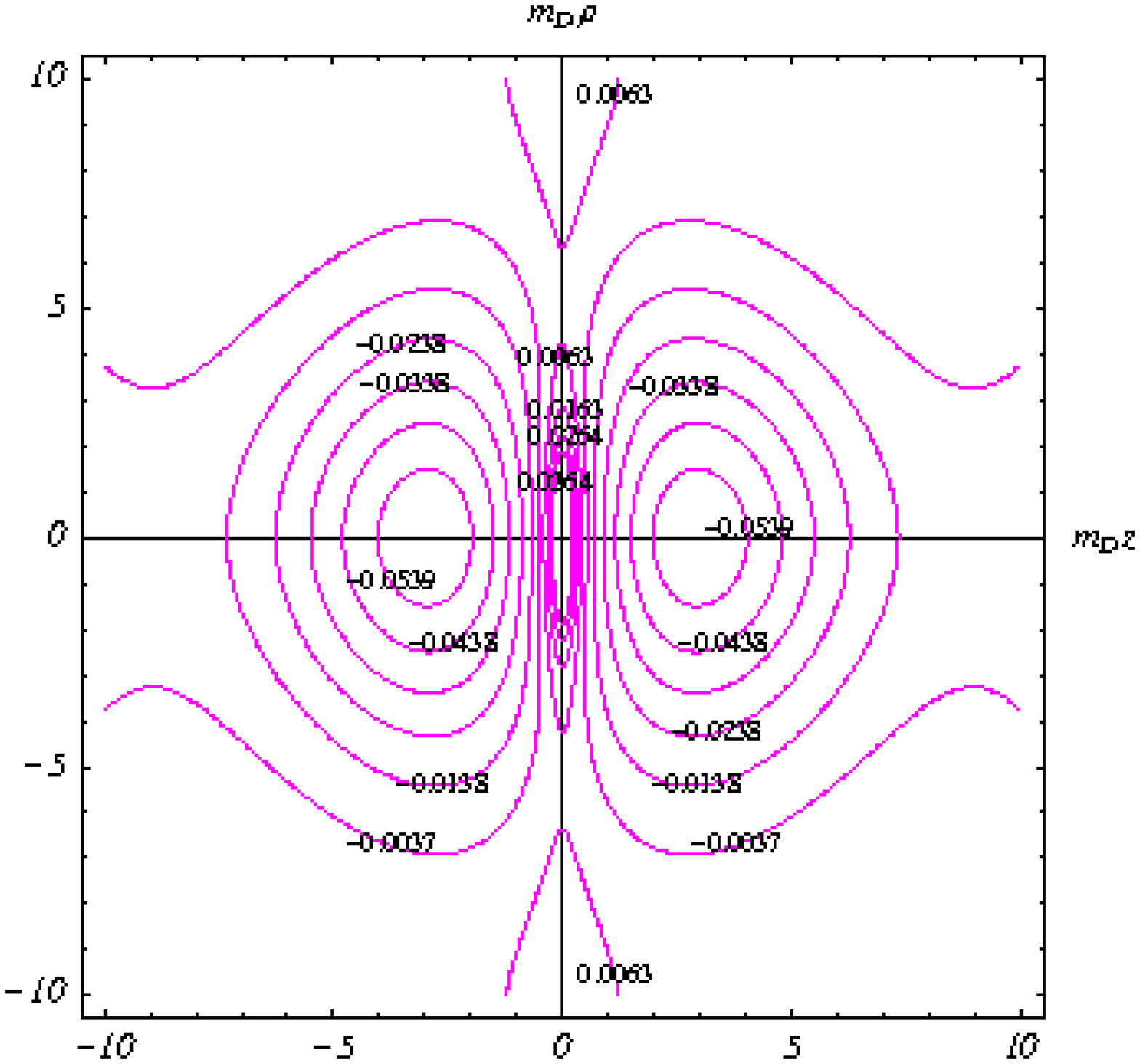}
\end{minipage}
%\vspace{0.8in}
\caption{Total contribution for $v=0.99c$}  
\label{fig:total_v99}
\end{figure}

With the increase of $v$, as we will see below, the relative strength of the
dipole potential grows strongly. For $v=0.99c$ Fig.~\ref{fig:long_v99} displays
the spatial distribution of the scaled longitudinal potential whereas
Fig.~\ref{fig:trans_v99} shows that of transverse part in the ($\rho - z$) plane. 
These plots clearly show that the magnitude of the spatial distribution of the
dipole potential due to both electric and magnetic interactions are enhanced
to a large extent in the ($\rho - z$) plane. The resulting potential distribution 
is shown in Fig.~\ref{fig:total_v99}. Apart from the usual features as 
discussed above for smaller $v$, it displays a new and important feature. 
A substantial repulsive interaction 
has grown in the transverse plane, {\it i.e.}, in the direction of $\rho$
but at $z=0$, becoming responsible for a vertical split of the minimum 
in the $(\rho -z)$ plane as can be seen from both the panels in 
Fig.~\ref{fig:total_v99}. 

\begin{figure}[!tbp]
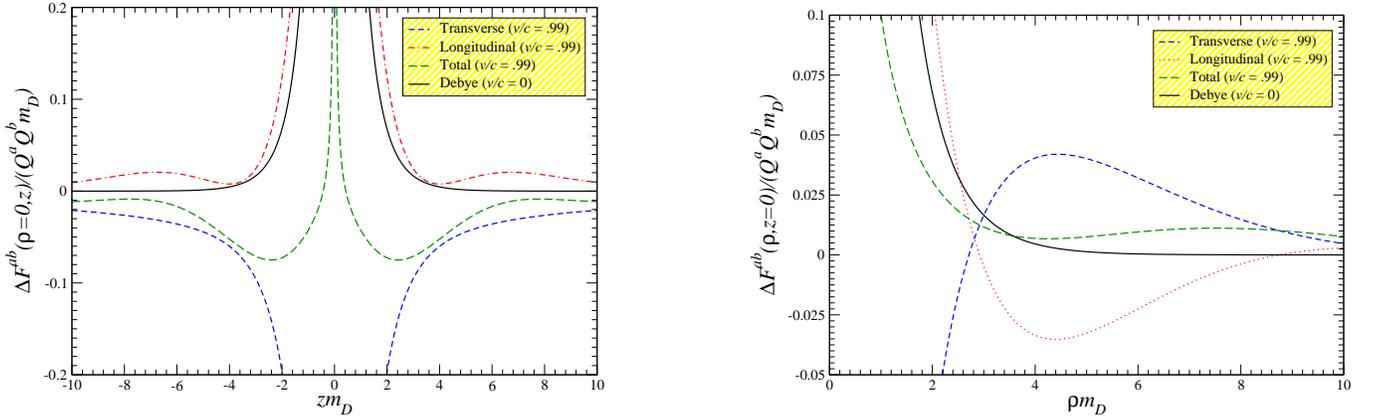

\vspace{.5cm}
%\begin{minipage}[h]{0.48\textwidth}
\begin{minipage}[t]{8cm}
\includegraphics[width=8cm,keepaspectratio]
{v99para_sc.eps}
\end{minipage}
\hfill
\begin{minipage}[t]{8cm}
\includegraphics[width=8cm,keepaspectratio]
{v99perp_sc.eps}
\end{minipage}
%\vspace{0.8in}
\caption{Same as Fig.~\ref{fig:specific_v55} but $v=0.99c$.}
\label{fig:specific_v99}
\end{figure}

Now, the detailed contributions of both the interactions can also be 
understood from both panels of Fig.~\ref{fig:specific_v99} where the 
dipole potentials for two special cases have been displayed. As shown in the
left panel of Fig.~\ref{fig:specific_v99} the depth of the negative 
minimum of the dipole potential along the $|z|$ direction increases and it 
position shifts towards  the centre indicating a faster fall off as 
well as a larger anisotropy. This is due to the fact that the magnitude of the 
magnetic interaction becomes dominant.
The form of the potential remains of the Lennard-Jones type.
On the other hand, the potential in the transverse direction ($z=0$) as shown 
in the right panel of Fig.~\ref{fig:specific_v99} is oscillatory as well as 
repulsive, which is again due to the dominant magnetic interaction 
in the transverse direction. So, with the increase of $v$ 
the magnetic interaction due to the transverse part of the response function 
plays a crucial role and its form deviates from the Lennard-Jones type.  
This could have important consequences on various binary states in the QGP.

In QCD the interaction between color charges in various channels is either 
attractive or repulsive. A quark and an antiquark correspond to the sum of 
irreducible
color representations~\footnote{The strength of the color 
interaction~\cite{Shuryak} can be calculated using $SU(3)$ color group.}: 
$\mathbf {3 \otimes {\bar 3}= {\bar 1} \oplus 8}$, where the interaction strength of
the color singlet representation is $-16/3$ (attractive) whereas that of 
the color octet channel is $2/3$ (repulsive).
Similarly, a two quark state corresponds to the sum of the
irreducible color representations: 
$\mathbf {3 \otimes  3 = {\bar 3} \oplus 6}$, where the antisymmetric color 
triplet is attractive ($-8/3$) giving rise to possible bound states~\cite{Shuryak}. 
The symmetric color sextet channel, on the other hand, is repulsive ($4/3$). 
Color bound states ({\it i.e.} diquarks) of partons at rest have been claimed 
by analyzing lattice data~\cite{Shuryak}.  The situation is different  
when partons are in motion. The dipole potentials along the 
direction of propagation and also normal to it have both attractive and 
repulsive parts, similar to the Lennard-Jones form. So, all the 
attractive channels or the repulsive channels in the static case get inverted
due to the two comoving partons constituting a dipole in the QGP. This could 
lead to dissociation of bound states (or to resonance states) as well 
formation of color bound states in the QGP. 

Recent lattice results~\cite{Datta} using maximum entropy method indicate
that charmonium states actually persist up to $2T_C$ and there are similar
evidence for mesonic bound states made of light quarks as well~\cite{Karsch}. 
Within our model such bound states as well other colored binary states in the QGP
can experience different potentials along the dipole direction and the direction 
normal to it. Along the direction of motion binary states which were  bound
in the QGP may become resonance states or dissociate beyond $T_C$. 
The temperature up to which they survive need further analysis of the bound 
states properties in detail, which is beyond the scope of this work. 
Similarly, those colored states which were not bound initially in the QGP, 
may transform into bound states.  There are some long distance 
correlations among partons in the QGP, which could indicate the 
appearance/disappearance of binary states in the QGP beyond $T_C$.
On the other hand, in the transverse direction the mesonic states 
as well as  colored bound binary states may dissociate for smaller velocity of
comoving partons whereas they may remain loosely bound if the comoving
partons are ultra-relativistic. One immediate consequence of this would be
the modification of the transverse momentum, $p_\bot$, dependence of 
$J/\psi$ spectra. Our analysis suggests that a QGP with moving partons beyond $T_C$ does not 
behave as a free gas of partons but a long range correlation may be present.

\section{Conclusions}

We have investigated the response of the QGP to a fast moving parton within
the HTL approximation valid in the high temperature limit. For velocities 
smaller than the phase velocity of the longitudinal plasma mode 
$v=\sqrt{3/5} c$ the wake potential corresponds to dynamical, anisotropic 
screening as discussed already before \cite{Mustafa05}. For larger velocities, 
however, we found, in contrast to previous investigations 
\cite{Mustafa05,Ruppert} that Mach cones and Cerenkov radiation appear in 
addition.
They have been identified by oscillations (sign flips) in the induced
charge density and also in the wake potential in the outward flow. 
This observation could be of phenomenological interest for the present
experimental programs in RHIC, where various approaches are being followed
to find Mach cones through the correlation between high $p_\bot$ secondary
particles in pseudorapidity and azimuthal space. The general structure of 
the wake potential in the direction of motion shows a long range correlation 
among the particles in colored plasma. The consequences for the formation of 
a possible liquid phase of the QGP \cite{Thoma_2005} should be investigated. 
In addition, the wake potential normal to the motion of the charged particle 
causes a transverse flow in the system which might contribute to the creation of 
instabilities \cite{Mrow} and an anomalous viscosity \cite{Asakawa}. 

Furthermore, we have investigated the dipole potential by comoving (anti)quarks 
in the QGP due to the appearance of a minimum in the wake potential of a moving
color charge, extending the work in Ref.\cite{Mustafa05}. In 
Ref.~\cite{Mustafa05} the dipole potential was obtained by averaging the one
body potential and the dipole potential was found to be dependent only on 
the electric interaction. In the present work we followed a proper treatment, as 
depicted in subsec.~\ref{dipole_pot}, and the dipole potential was found to depend on both 
electric and magnetic interactions. As discussed both interactions play a 
crucial role in determining the nature of the dipole potential. Based on this  
we analyzed the various possibilities, {\it viz.}, the appearance/disappearance
of bound states in the QGP by discussing the effects of Ampere's law and the 
nature of electric and magnetic forces due to dynamical screening.

\section*{Acknowledgment} 
MGM is thankful to Subhasis Chattopadhyay for various useful discussion.

\end{document}